\documentclass[12pt]{article}

\usepackage{latexsym}

\font\tenmsbm=msbm10 scaled 1200
\font\sevenmsbm=msbm9
\newfam\msbmfam
\textfont\msbmfam=\tenmsbm
\scriptfont\msbmfam=\sevenmsbm

\makeatletter
\@addtoreset{equation}{section}
\makeatother

\newcommand{\eref}[1]{(\ref{#1})}

\def\beq{\begin{equation}}
\def\eeq{\end{equation}}
\def\bea{\begin{eqnarray}}
\def\eea{\end{eqnarray}}
\def\bet{\begin{tabular}}
\def\eet{\end{tabular}}
\def\pa{\partial}

\def\ve{\varepsilon}
\def\qua{\quadratello}

\catcode`@=11
\def\lsim{\mathchoice
  {\mathrel{\lower.8ex\hbox{$\displaystyle\buildrel<\over\sim$}}}
  {\mathrel{\lower.8ex\hbox{$\textstyle\buildrel<\over\sim$}}}
  {\mathrel{\lower.8ex\hbox{$\scriptstyle\buildrel<\over\sim$}}}
  {\mathrel{\lower.8ex\hbox{$\scriptscriptstyle\buildrel<\over\sim$}}} }
\def\gsim{\mathchoice
  {\mathrel{\lower.8ex\hbox{$\displaystyle\buildrel>\over\sim$}}}
  {\mathrel{\lower.8ex\hbox{$\textstyle\buildrel>\over\sim$}}}
  {\mathrel{\lower.8ex\hbox{$\scriptstyle\buildrel>\over\sim$}}}
  {\mathrel{\lower.8ex\hbox{$\scriptscriptstyle\buildrel>\over\sim$}}} }
\def\croce{\displaystyle / \kern-0.2truecm\hbox{$\backslash$}}
\def\lqua{\lower4pt\hbox{\kern5pt\hbox{$\sim$}}\raise1pt
\hbox{\kern-8pt\hbox{$<$}}~}
\def\gqua{\lower4pt\hbox{\kern5pt\hbox{$\sim$}}\raise1pt
\hbox{\kern-8pt\hbox{$>$}}~}
\def\mma{\lower1pt\hbox{\kern5pt\hbox{$\scriptstyle <$}}\raise2pt
\hbox{\kern-7pt\hbox{$\scriptstyle >$}}~}
\def\mmb{\lower1pt\hbox{\kern5pt\hbox{$\scriptstyle >$}}\raise2pt
\hbox{\kern-7pt\hbox{$\scriptstyle <$}}~}
\def\mmc{\lower4pt\hbox{\kern5pt\hbox{$<$}}\raise1pt
\hbox{\kern-8pt\hbox{$>$}}~}
\def\mmd{\lower4pt\hbox{\kern5pt\hbox{$>$}}\raise1pt
\hbox{\kern-8pt\hbox{$<$}}~}
\def\lsu{\raise4pt\hbox{\kern5pt\hbox{$\sim$}}\lower1pt
\hbox{\kern-8pt\hbox{$<$}}~}
\def\gsu{\raise4pt\hbox{\kern5pt\hbox{$\sim$}}\lower1pt
\hbox{\kern-8pt\hbox{$>$}}~}
\def\croce{\displaystyle / \kern-0.2truecm\hbox{$\backslash$}}
\def\ali{\hbox{A \kern-.9em\raise1.7ex\hbox{$\scriptstyle \circ$}}}
\def\2frecce{\hbox{\lower 0.3ex\hbox{$\leftarrow$} 
\hbox{\kern-1.3em\raise 0.3ex\hbox{$\rightarrow$}}}}
\def\quad@rato#1#2{{\vcenter{\vbox{
        \hrule height#2pt
        \hbox{\vrule width#2pt height#1pt \kern#1pt \vrule width#2pt}
        \hrule height#2pt} }}}
\def\quadratello{\mathchoice
\quad@rato5{.5}\quad@rato5{.5}\quad@rato{3.5}{.35}\quad@rato{2.5}{.25} }

\begin{document}

\begin{titlepage}

\begin{flushright}
Preprint DFPD 00/TH/32\\
July 2000\\
\end{flushright}

\vspace{0.5truecm}

\begin{center}

{\Large \bf Interacting branes, dual branes, and dyonic branes:}
\par\vspace{0.4cm}
{\Large \bf a unifying lagrangian approach in $D$ dimensions}

\vspace{1.5cm}

{K. Lechner\footnote{kurt.lechner@pd.infn.it}  and
{P.A. Marchetti}\footnote{pieralberto.marchetti@pd.infn.it}} 
\vspace{2cm}

{ \it Dipartimento di Fisica, Universit\`a degli Studi di 
Padova,

\smallskip

and

\smallskip

Istituto Nazionale di Fisica Nucleare, Sezione di Padova, 

Via F. Marzolo, 8, 35131 Padova, Italia
}

\vspace{1cm}

\begin{abstract}
\vspace{0.5cm}

This paper presents a general covariant lagrangian framework for
the dynamics of a system of closed $n$--branes and dual 
$(D-n-4)$--branes in $D$ dimensions, interacting  with a dynamical
$(n+1)$--form gauge potential. The framework proves sufficiently
general to include also a coupling of the branes to (the bosonic
sector of) a dynamical supergravity theory.
We provide a manifestly Lorentz--invariant
and $S$--duality symmetric Lagrangian, involving the $(n+1)$--form 
gauge potential and its dual $(D-n-3)$--form gauge potential 
in a symmetric way. The corresponding action depends on generalized 
Dirac--strings. The requirement of string--independence of the 
action leads to Dirac--Schwinger quantization conditions
for the charges of branes and dual branes, but produces also additional
constraints on the possible interactions.
It turns out that a system of interacting dyonic branes 
admits two quantum mechanically inequivalent formulations, involving 
inequivalent quantization conditions. 
Asymmetric formulations involving only a single
vector potential are also given. For the special cases of 
dyonic branes in even dimensions known results are easily recovered. 
As a relevant application of the method we write an effective action which
implements the inflow anomaly cancellation mechanism for interacting 
heterotic strings and five--branes in $D=10$. A consistent realization of 
this mechanism requires, in fact, dynamical $p$--form potentials and a 
systematic introduction of Dirac--strings.

\end{abstract}

\end{center}
\vskip 0.5truecm 
\noindent PACS: 11.15.-q, 11.10.Kk, 11.30.Cp; Keywords: branes, duality,
covariance.
\end{titlepage}

\newpage

\baselineskip 6 mm


\section{Introduction}

The unification of the known consistent string theories in
a fundamental conjectured $M$--theory in eleven dimensions relies
heavily on the discovery of a set of new extended objects, 
so called $n$--branes. Since a basic concept in this unification 
is duality, electrically charged $n$--branes in $D$ dimensions
are naturally accompanied by magnetically charged dual $(D-n-4)$--branes.
The quantum consistency of a theory in which both types of excitations are
present relies on a quantization condition for the product 
of their electric and magnetic charges. 

The low energy dynamics of $n$--branes in string theory
is usually described by a Green--Schwarz sigma model in a (super)gravity 
background, where the background fields, to cope with $\kappa$--symmetry,
have to satisfy their sourceless {\it free} equations of motion: 
one neglects the effect of the backcoupling of the branes on the dynamics 
of the supergravity fields themselves. Since until now a 
$\kappa$--invariant action describing the dynamics of an {\it interacting} 
brane--supergravity system is not known, $\kappa$--symmetry can not be used
to get information on the nature of this backcoupling.

The present paper opens a different channel for the investigation
of the dynamics of such an interacting system, in the absence of fermions,
in terms of an action principle.
The key--ingredients of the strategy are represented by a systematic 
introduction of Dirac--strings, i.e. of surfaces whose boundaries are  
the brane worldvolumes, and of a related manifestly Lorentz invariant action 
principle describing the interaction between branes and gauge potentials.
The introduction of Dirac--strings is, indeed,  unavoidable in the following
relevant cases: first, when branes and dual branes are simultaneously 
present in a theory, a well known feature in the case of charges and monopoles
in four dimensions; second, when the dynamics of the supergravity theory 
can not be expressed in terms of the gauge--potential alone which is 
electrically coupled to the $n$--brane, but requires also the introduction
of its dual potential, \cite{Dual}. A typical example of the second
scenario is represented by the
$M5$--brane in eleven dimensions which couples electrically to the 
six--form potential
$A_6$ of $N=1$, $D=11$ supergravity. Since the invariant curvature
for $A_6$, given by 
$$ 
H_7=dA_6+{1\over2}A_3\,dA_3,
$$
involves its magnetic dual $A_3$, the supergravity dynamics can not be
expressed in terms of $A_6$ alone. This feature requires a {\it magnetic}
coupling of the $M5$--brane to $A_3$, and hence the introduction of 
Dirac--strings. A similar situation occurs for example for $NS5$--branes 
in $D=10$, $IIB$ supergravity and $D4$--branes in $IIA$--supergravity.

An appropriate treatment of these situations requires therefore a 
framework in which potentials and dual potentials appear in a symmetric
way. An efficient approach which realizes such a framework
is given by the PST--method \cite{PST}, thanks to its manifest invariance
and to its compatibility with all known symmetries. It will therefore be 
used as a convenient technical ingredient in the present paper.

On the other hand, the presence of Dirac--strings introduces a new
consistency requirement on the dynamics: physics has to be independent
of the particular choice of the Dirac--string, i.e. the string has to be
unobservable. This requirement imposes a quantization condition for 
the charges, but constrains also the nature of non--minimal interactions
between supergravity and $n$--branes. In this sense Dirac--string 
independence, as we will see, determines heavily the structure of the 
backcoupling.

The aim of this paper is to settle a general covariant and flexible 
lagrangian framework for the description of the dynamics of branes and dual
branes interacting with gauge potentials. In the first larger part 
we will concentrate mainly on the general properties of this 
system: Lorentz--invariance, effective actions, Dirac quantization 
conditions, $S$--duality. Particular attention will be paid to dyons
in $4K$ and $4K+2$ dimensions, section six. As has been noted some time ago
\cite{SCHW2,ZW2} the theories of dyons comes in in two different physically 
{\it inequivalent} versions, which are governed by inequivalent 
Dirac--quantization conditions. For each of these theories we will
give two {\it equivalent formulations}; a Schwinger--like
formulation in terms of a single vector potential, and a PST--formulation
in terms of two vector potentials. 
It is understood that some of the results of this section
have been already derived in \cite{DESER}, in a non--covariant framework.

Sections seven and eight are devoted to non--minimal couplings. Particularly
interesting are the couplings induced by the inflow anomaly cancellation
mechanism. We will work out in detail a simple, but non trivial, example
represented by 
heterotic strings and five--branes interacting with $N=1$, $D=10$
supergravity. The non trivial feature is in this case represented by
a modification of the equation of motion for the $B_2$--potential,
induced by the residual anomaly on the five--brane.

Further applications of our framework to the inflow mechanism for
$M2$-- and $M5$--branes in eleven dimensions 
and $NS5$--branes in $IIA$--supergravity are deferred to a future
publication \cite{LMT}.

\section{Algebra of differential forms} 

A natural framework for a theory of branes interacting with gauge
potentials of higher rank is provided by the language of differential 
forms. Apart from the notational convenience of this language -- one
avoids writing indices -- it fits also naturally with the 
extension of Poincar\`e--duality to ``singular" $p$--form currents, 
a concept used widely in the text. In this paper space--time is
assumed to be topologically trivial so that, in particular, one can make
free use of the Poincar\`e lemma, also for currents.

We summarize here our basic conventions regarding  differential forms in
$D$ dimensions. Our $D$--dimensional space--time is endowed with a 
metric tensor $G_{\mu\nu}(x)$, and the flat metric carries 
Minkowsky signature $\eta^{\mu\nu}=(1,-1,\cdots,-1)$; the Levi--Civita 
tensor is characterized by $\ve^{01\cdots D-1}=1$. A $p$--form is 
decomposed along its coordinate basis according to 
$$
\Phi={1\over p!}\,dx^{\mu_1}\cdots dx^{\mu_p}\,\Phi_{\mu_p\cdots \mu_1}.
$$
The Hodge--dual of $\Phi$, the $(D-p)$--form  
$$
*\Phi={1\over (D-p)!}\,
dx^{\mu_1}\cdots dx^{\mu_{D-p}}\,(*\Phi)_{\mu_{D-p}\cdots \mu_1},
$$
is defined by 
\beq
\label{hodge}
(*\Phi)_{\mu_1\cdots\mu_{D-p}}= (-)^{{1\over2}p(p+1)}\,{1\over p!}\,
G_{\mu_1\alpha_1}\cdots G_{\mu_{D-p}\alpha_{D-p}}
{\varepsilon^{\alpha_1\cdots\alpha_{D-p}\nu_1\cdots \nu_p}
\over \sqrt{G}}\,\Phi_{\nu_1\cdots \nu_p}.
\eeq
The sign in the definition of the Hodge--dual is a matter of convention
\footnote{The sign chosen in \eref{hodge} differs from the one used
in \cite{LM}; this leads to some convention depending sign differences between 
the present paper and \cite{LM}, when one specializes to the case $D=4$.}.
The advantage of the choice \eref{hodge} is that the $*$--operator squares on a 
$p$--form to a sign which is {\it independent} of $p$ and depends only
on the dimension:
$$
*^2= \eta \equiv (-1)^{{1\over2}D(D-1)+1}.
$$
In even dimensions, $D=2N$, one has $\eta=(-1)^{N+1}$ and this implies 
the existence of self--dual (and antiself--dual) $N$--forms only for
odd $N$, as is well known. 

The differential $d$ acts on a form from the right, i.e.
$d\left(\Phi_p\Phi_q\right)=\Phi_p d\Phi_q+(-)^q(d\Phi_p)\Phi_q$, and
$d^2=0$. 
Forms in the image of $d$ are called exact and forms in the kernel of $d$ are 
called closed. Since we work in a topologically trivial space--time
all closed forms are also exact.
The product between forms will always be an exterior (wedge) product and 
the wedge symbol $\wedge$ will be omitted. The codifferential is defined by 
$\delta = *d*$ and lowers the rank of a form by 1. The D'Alambertian
$\qua=D_\mu G^{\mu\nu}D_\nu$ admits on a $p$--form the standard decomposition
$$
\qua=\eta (-)^D(d\delta +\delta d).
$$

For a generic one--form $U=dx^\mu U_\mu$ we indicate with $i_U$ the 
interior product between the conjugated vector field $U^\mu\pa_\mu$
and a $p$--form, defined as
$$
i_U \Phi={1\over (p-1)!}\,dx^{\mu_1}\cdots dx^{\mu_{p-1}}U^{\mu_p}
\,\Phi_{\mu_p\cdots \mu_1}.
$$
The operator $i_U$ lowers the rank of a form by 1 and satisfies
the same distribution law as $d$. If $U^\mu$ is nowhere
lightlike a differential form can be uniquely decomposed in a component along
$U$ and a component orthogonal to $U$. The relevant operatorial identities 
on $p$--forms are
\bea
1&=&(-)^{p+1}\,{1\over U^2}\,\left(Ui_U+i_UU\right)
=(-)^{p+1}\,{1\over U^2}\,\left(Ui_U+(-)^D\eta *Ui_U*\right)\nonumber\\
{\phantom{x}}*&=& (-)^{p+1}\,{1\over U^2}\left(*Ui_U+(-)^DUi_U*\right)
\nonumber\\
U*&=&(-)^p*i_U.
\label{ident}
\eea

\section{A covariant action for external brane--sources} 

In a $D$--dimensional space--time $n$--branes are dual to 
$(D-n-4)$--branes. $n$--branes carry a natural electric coupling 
to an $(n+1)$--form gauge potential and a magnetic coupling to the
dual $(D-n-3)$--form gauge potential; for the dual branes the types 
of the couplings are reversed. In this picture, which uses two gauge 
potentials, the dynamics of the gauge degrees of freedom is governed
by the Hodge--duality relation between the field strengths associated to 
the two potentials.

In this section we want describe the dynamics of the gauge degrees of 
freedom, in the presence of external conserved brane--currents, in terms
of a covariant action. The equations of motion which describe this system
can be written in the form of generalized Maxwell's equations as 
\bea
dF^1&=&J^1 \label{eq1}\\
dF^2&=&J^2 \label{eq2}\\
F^1=*F^2&\Leftrightarrow& F^2=*\eta F^1.
\label{dual}
\eea
The curvatures  $F^I$ ($I=1,2$) are forms of rank $p^I$ with
\bea
p^1&=&n+2\\
p^2&=&D-n-2,
\eea
and \eref{eq1} and \eref{eq2} imply that the currents, $(p^I+1)$--forms, are
conserved
\beq
\label{cons}
dJ^I=0.
\eeq
For the time being they can also correspond to continuous brane distributions
and are not required to be $\delta$--functions on a brane 
worldvolume. The conservation equations imply the existence of forms $C^I$
such that
\beq
J^I=dC^I.
\label{defc}
\eeq
They are again forms of rank $p^I$. Eqs. \eref{eq1} and \eref{eq2} can then 
be solved introducing potentials $A^I$, $(p^I-1)$--forms, such that
\beq
\label{def}
F^I=dA^I+C^I.   
\eeq
With these definitions of potentials and curvatures the dynamics of the
gauge degrees of freedom is described by the duality relation \eref{dual},
which is promoted to an equation of motion.

A Lorentz--invariant action which gives rise to \eref{dual} can be 
constructed using a method introduced by Pasti, Sorokin and Tonin (PST)
\cite{PST}. It requires the introduction of a single scalar auxiliary
field $a(x)$ and of the one--form 
$$
v={da\over \sqrt{-G^{\mu\nu}\pa_\mu a\, \pa_\nu a}}\equiv dx^\mu v_\mu,
$$
whose components satisfy $v^2=v^\mu v_\mu=-1$. In this framework
the problem of Lorentz--invariance is converted in the requirement
that the field $a$ becomes auxiliary (non propagating). In the 
PST--approach the auxiliary nature of $a$ relies on the particular 
symmetries of the PST--action; one of them allows to fix $a$
to an arbitrary value. In this sense the PST--symmetries represent nothing 
else then the requirement of Lorentz--invariance, and they constrain the 
form of the action in any dimensions. 

The action can be conveniently written in terms of the $(p^I-1)$--forms 
\bea
f^1 &\equiv& i_v(F^1-*F^2)\\
f^2 &\equiv& i_v(F^2-*\eta F^1),
\eea
which parametrize the decomposition of the duality--relation \eref{dual},
associated to $v$ (see \eref{ident} with $U=v$):
\beq
F^1-*F^2=(-)^n\left(vf^1+(-)^{D+1} *vf^2\right).
\label{dec}
\eeq
We write the PST--action, as the integral of a $D$--form, in three 
different ways,
\bea
I_0[A,C,a]&=&\int\left[{1\over2}\left(F^1*F^1+(-)^n f^1*f^1\right)
             -\eta\left(dA^1C^2+{1\over2}C^1C^2\right)\right]
            \nonumber\\         
&=& \eta(-)^{nD+n+1}
\int\left[{1\over2}\left(F^2*F^2+(-)^{D+n}f^2*f^2\right)
             -\left(dA^2C^1+{1\over2}C^2C^1\right)\right]\nonumber\\   
&=&{1\over2}\int\left[ F\,{\cal P}(v)\,F +\eta
\left(C^1dA^2-dA^1C^2\right)\right]. 
\label{PST}
\eea
It is understood that the dynamical variables are $A^1$, $A^2$ and $a$. 
The curvatures are defined as in \eref{def} and with $F$ we indicate
the ordered couple $(F^1,F^2)$, with form degrees $(p^1,p^2)$. The operator
${\cal P}(v)$ sends a couple of $(p^1,p^2)$--forms in a couple of
$(p^2,p^1)$--forms, such that $F\,{\cal P}(v)\,F$ is a $D$--form. It is
defined as
\beq
{\cal P}(v)=
\left(\matrix{(-1)^{D+n}& 0\cr 
                       0&\eta(-)^{nD+1}\cr}\right)vi_v*
+\left(\matrix{0&\eta(-)^{D+n+1}\cr 
    \eta(-)^{nD}&0\cr}\right)vi_v
+{1\over2}
  \left(\matrix{0&\eta\cr 
\eta(-)^{nD+n+1}&0\cr}\right)
\eeq
This operator is symmetric in the sense that for two couples $F$ and
$G$ of $(p^1,p^2)$--forms one has $F\,{\cal P}(v)\,G=G\,{\cal P}(v)\,F$.
  
The three different expressions of the action can be obtained making 
repeated use of the identities \eref{ident} and of $*^2=\eta$.
The first line of \eref{PST} privileges the potential $A^1$. If one drops
the term $f^1*f^1$, proportional to the square of the duality relation
\eref{dual}, then one obtains a formulation in terms of only the single 
potential $A^1$. The corresponding action generalizes the one given
by Schwinger \cite{SCHW2} for the dynamics of dyons in four dimensional 
electrodynamics.

The second line privileges the potential $A^2$ and can be
obtained from the first line with the formal substitutions 
$1\leftrightarrow2$, $n\leftrightarrow D-n-4$. The additional factor
of $\eta$ comes from the $\eta$ which appears in the dual duality relation
$F^2-\eta* F^1=0$ 
\footnote{For notational convenience the overall sign in $I_0$ has been 
chosen as $+1$. To obtain a positive definite energy the formulae in
\eref{PST} have to be supplied by an additional overall sign given
by $\gamma(n)\equiv\eta (-1)^{ {1\over2}n(n+1)+D+n+1}$. This ensures 
positiveness of the term $F^1*F^1$ in the first line. The positiveness of the 
term $F^2*F^2$ in the second line is then guaranteed by the identity  
$\gamma(n)\eta(-1)^{nD+n+1}=\gamma(D-n-4)$.}. In the third 
line the potentials $A^1$ and $A^2$ appear in a symmetric way.
This form will appear 
very useful for displaying the $S$--duality properties of the action and
for establishing the duality--symmetry groups in the case of dyonic branes.

As we will see, at the quantum level
the theory of interacting dyons comes in in two {\it inequivalent} versions,
while the theory of interacting branes and dual branes admits, actually, 
only a  single version. To display from the very beginning
the difference between the two versions, which we call respectively 
``symmetric" and ``asymmetric", we illustrate them  here also for 
the case of branes and dual branes.
The action we wrote above corresponds to the ``symmetric" theory, in the sense 
that it is invariant under the formal interchange $1\leftrightarrow2$,
as explained above. The ``asymmetric" theory is obtained by adding to, or 
subtracting from, $I_0$ the term ${1\over2}\eta\int C^1C^2$. In the first
case one cancels in the first line in \eref{PST} the last term, in the 
second case one cancels the last term in the second line. In the following 
we will show that the two choices, obtained by subtracting or adding
this term, lead in {\it any} case to equivalent quantum theories. For 
definiteness we choose here for the action of the asymmetric theory
\beq
\widetilde I_0\equiv I_0 +{1\over2}\eta\int C^1C^2.
\eeq
Since the added term is independent from $A^I$,
it is clear that $I_0$ and $\widetilde I_0$ lead to the same equations
of motion for the gauge potentials. We will also see that in the case of
{\it dynamical} branes the equations of motion for the branes will be 
identical too. The inequivalence between the symmetric and asymmetric 
theories will arise in the {\it quantum} theory of dyons, as we will see 
below; in particular, for the two theories we will obtain different 
quantization conditions for the charges.
    
As we mentioned above, a one potential formulation can be obtained by 
dropping in the first line in \eref{PST} the term $f_1*f_1$. For the 
symmetric theory this leads to the Schwinger--like action, in terms of $A_1$, 
$$
I_0 \rightarrow I_{Schwinger}=
\int\left[{1\over2}F^1*F^1 -\eta dA^1C^2-{1\over2}\eta\int C^1C^2\right].
$$
For the asymmetric theory one has 
$$
\widetilde I_0 \rightarrow \widetilde I_{Schwinger}=
\int\left[{1\over2}F^1*F^1 -\eta dA^1C^2\right].
$$
$F^1$ satisfies the Bianchi identity $dF^1=J^1$ and the equation of motion
for $A^1$ is $d(*\eta F^1)=J^2$. These are just the generalized Maxwell
equations \eref{eq1}--\eref{dual}. 
Whenever there is no need to introduce both potentials, this form of the
action for the theory is the most convenient and simple one.
We would like to stress that the formulations in terms of a single vector
potential and the ones in terms of two vector potentials a la PST are
completely equivalent. When the dynamics is described by the simple
set \eref{eq1}--\eref{dual} both options are reliable: the PST--formulation
appears more convenient for the description of the symmetric theory 
(manifest duality invariance), while the Schwinger--formulation appears
more convenient for the asymmetric theory. When the dynamics is more
complicated (e.g. self--interactions of the potential, chiral bosons, 
presence of Chern--Simons terms) the PST--formulations might be unavoidable.
For definiteness in what follows we use the PST--formulation.

For completeness we illustrate now briefly the structure of the 
PST--symmetries. 
All signs and relative coefficients in the action $I_0$ are indeed 
uniquely determined by those symmetries. The corresponding transformations 
are given 
by
\bea
\label{1}\delta A^I&=& d\Lambda^I \\
\label{2}\delta A^I&=& \Phi^I\, da\\
\label{3}\delta A^I&=& - {\varphi\over \sqrt{-(\pa a)^2}}\,f^I, \quad 
            \delta a= \varphi.
\eea
The transformation parameters are the $(p^I-2)$--forms  $\Lambda^I$ and 
$\Phi^I$, and the single scalar $\varphi$. We relegate the proof of the 
invariance of $I_0$ under these transformations to the appendix. 
The transformations
in \eref{1} are just ordinary gauge transformations for the 
potentials $A^I$, and \eref{3} states that the field $a$ is a non propagating
auxiliary field which can be shifted to any value.   
The transformations \eref{2} allow to reduce the second order equation
of motion for the gauge fields to the first order duality relation 
\eref{dual}. To see this we write the equations of motion for $A$ and $a$,
derived in the appendix, 
\bea
\label{4}
d\,(vf^I)&=&0\\
\label{5}
d\left({v\over \sqrt{-(\pa a)^2}}\,f^1f^2\right)&=&0.
\eea
The equation of motion for $a$ \eref{5} is a consequence of the 
$A^I$--equations \eref{4} and contains no dynamical information.
The general solution of \eref{4} is $vf^I=da\,d{\widetilde{\Phi^I}}$, 
for some $(p^I-2)$--forms ${\widetilde{\Phi^I}}$; it can then be seen
\cite{PST} that through a transformation 
\eref{2}, with $\Phi^I={\widetilde{\Phi^I}}$, one can set $f^I=0$. But, 
due to the identity \eref{dec} this is equivalent to \eref{dual}.
Identical conclusions hold also for the action $\widetilde I_0$ of the
asymmetric theory. 

Until now we ignored an ambiguity which is associated to the choice of the 
$C$--fields. From their definition \eref{defc} it is clear that they
are defined only modulo exact forms. Since the curvatures should not be 
affected by this ambiguity, one has to consider the combined finite
transformations
\beq
\Delta C^I=dW^I, \quad \Delta A^I=-W^I,
\label{change}
\eeq
where the $W^I$ are $(p^I-1)$--forms. Whereas the equations of motion
for the gauge fields \eref{4} are invariant under these transformations 
the action is not. We call the transformation of the action under 
\eref{change} ``Dirac--anomaly", for a reason that will become clear in 
the next section. For the symmetric and asymmetric theories respectively,
the finite Dirac--anomaly (i.e. for finite variations) can be easily 
computed to be 
\footnote{Considering e.g. the third line in \eref{PST} one sees that the 
first term is trivially invariant, while in the second and third terms
only the variations of $A^I$ contribute.}
\bea
\label{diracs}
A_D\equiv \Delta I_0&=& {1\over 2}\eta(-)^{D+n+1}\int
                       \left(W^1J^2+J^1 W^2\right)\\
\widetilde A_D\equiv \Delta \widetilde I_0&=& \eta(-)^{D+n+1}
                         \int W^1 J^2.
\label{diraca}
\eea
In the presence of dynamical branes the transformations \eref{change} will 
correspond to a generalized change of Dirac--strings, and the vanishing of
the (exponentiated) Dirac--anomaly will trigger the consistency of the 
quantum dynamics of these extended objects and impose quantization 
conditions for the charges. The difference between the formulae 
\eref{diracs} and \eref{diraca} reflects the inequivalence 
between the symmetric and asymmetric theories, mentioned above. 

The Dirac--anomalies associated to the Schwinger--like actions are also
given by \eref{diracs}, \eref{diraca}; this is a trivial consequence of the 
equivalence between the PST-- and Schwinger--formulations.

One more word is in order for what concerns the identification of the
observable degrees of freedom carried by the gauge fields. A look on the 
structure
of the PST--symmetries, more precisely on \eref{2}, shows that the 
curvatures $F^I$ are not invariant objects, even if the equations
of motion \eref{4} are satisfied; therefore, they do not correspond
to observable quantities. The observable curvatures, which can indeed 
be seen
to be invariant under \eref{1}--\eref{3} if \eref{4} holds, are given
by
\bea
K^1&=&F^1+(-)^{n+1}vf^1\\
K^2&=&F^2+(-)^{D+n+1}vf^2.
\eea
They are also invariant under \eref{change} and, moreover, they satisfy 
identically the duality relation 
$$
K^1=*K^2.
$$
Once the symmetry \eref{2} is fixed as above, i.e. $f^I=0$, one has 
$K^I=F^I$. Notice that in terms of the $K^I$--tensors the equations 
of motion \eref{4} can be written also as 
\beq
dK^I=J^I.
\label{dualk}
\eeq
Therefore, the $K^I$'s satisfy the same equations as the $F^I$'s, i.e.
\eref{eq1}--\eref{dual}. For the $F^I$'s \eref{eq1} and 
\eref{eq2} are identities and the duality relation is an equation of
motion. For the $K^I$'s the situation is reversed: they satisfy the
duality relation identically and \eref{dualk} is their equation of motion.

To illustrate the implementation of Lorentz--invariance in the present
framework we derive now the current--current effective action. The 
knowledge of this effective action will display also more clearly the
difference between symmetric and asymmetric theories. For the symmetric
theory it is defined by the (normalized) functional integral over 
$A^I$ and $a$ of the exponentiated classical action,
\beq
e^{i\Gamma[C]}=\int \{ {\cal D} A\}\,\{ {\cal D}a\}\,
               \,e^{iI_0[A,C,a]}\,\Bigg/\,
\int \{ {\cal D} A\}\,\{ {\cal D}a\}\,
               \,e^{iI_0[A,0,a]}.
\label{defg}
\eeq
Appropriate gauge fixings of the PST--symmetries are understood, in 
particular the insertion of a $\delta$--function $\delta(a-a_0)$
for the symmetry \eref{3}, where $a_0(x)$ is an arbitrary scalar
gauge fixing function.  
The requirement of Lorentz--invariance translates here in the requirement
that $\Gamma[C]$ is independent of the (unphysical) function $a_0$.
But this is ensured by the PST--symmetry \eref{3} of the action $I_0$.
An explicit evaluation of the right hand side of \eref{defg} 
(a sketch of the computation is given in the appendix) leads indeed to the
$a_0$--independent result
\beq
\label{gamma}
\Gamma[C]=(-)^D{1\over2}\int\left(J {*\over \qua}\,{\cal M}\,J 
          + J {\delta\over\qua}\,{\cal N}\,C\right).
\eeq
Here $J=(J^1,J^2)$, $C=(C^1,C^2)$ and the $2\times 2$ matrices ${\cal M}$
and ${\cal N}$ are respectively diagonal and off--diagonal,
\bea 
{\cal M}&=&\left(\matrix
        {(-1)^{D+n}&0\cr 
         0& \eta(-1)^{nD+1}\cr}\right)\\
{\cal N}&=&\left(\matrix
{0&(-1)^{D+n+1}\cr 
(-1)^{nD}&0\cr}\right).
\eea
For the asymmetric theory the effective action becomes
\bea
\nonumber
\widetilde \Gamma[C]&=&  (-)^D \int\left({1\over2}J {*\over \qua}\,{\cal M}\,J 
         + (-)^{nD} J^2 {\delta\over\qua}C^1\right)\\
\label{gammaas}  &=&\Gamma[C]+{1\over2}\eta\int C^1C^2.
\eea
By construction the difference between $\widetilde \Gamma$ and 
$\Gamma$ equals the difference between $\widetilde I_0$ and $ I_0$.
The diagonal terms in $\Gamma$ and $\widetilde \Gamma$, which involve
only the currents $J$, coincide and represent the Coulomb--like 
electric--electric and magnetic--magnetic interactions.
The mixed terms, representing the electric--magnetic 
interactions, involve also the ``strings" $C$ and differ by the term 
${1\over2}\eta\int C^1C^2$. The Dirac--anomalies carried by the
effective actions, defined as their variations under $C^I\rightarrow 
C^I+dW^I$, are, by construction, given again by \eref{diracs} and
\eref{diraca}. The technical tool involved in checking these assertions on the
effective actions is the standard Hodge decomposition of the D'Alambertian, 
in the form
$$
1=\eta (-)^D(d\delta+\delta d)\cdot {1\over \qua}.
$$
More explicit expressions for the effective actions will be given 
for the particular cases considered below.

We would like to stress that the effective actions obtained from 
the Schwinger formulations, with a single vector potential, coincide with 
the ones given above; this is again a consequence of the equivalence of 
the two formulations.

\section{Dynamical $n$--branes and dual $(D-n-4)$--branes}

Until now we took the currents $J^I$ as arbitrary conserved external
currents. The physically interesting case corresponds to currents which
are associated to a certain number of extended charged  objects, i.e.
charged branes. The aim of the present section is to describe the 
interaction of the potentials $A^I$ with a set of dynamical branes and
dual branes in terms of a covariant action. The essential new feature 
of this situation is represented by the requirement of the vanishing of 
the (exponentiated) Dirac--anomaly, leading eventually to charge 
quantization. The analysis of the Dirac--anomaly can be carried out most
easily using the concept of Poincar\`e--duality between hypersurfaces 
and ``$p$--currents". So it will be briefly reviewed below.

We consider a system of a certain number $N_2$ of closed $n$--branes,
with charges $e^2_r$ and tensions $T^2_r$ ($r=1,\cdots,N_2)$,
and of $N_1$ closed dual $(D-n-4)$--branes, with charges $e^1_r$ and tensions 
$T^1_r$ ($r=1,\cdots,N_1)$. 

During time evolution an $n$--brane sweeps out a boundaryless 
$(n+1)$--dimensional
worldvolume parameterized by $y^\mu(\sigma^i)$, where $i=(0,1\cdots,n)$
and $\sigma^0$ indicates the evolution parameter. The dynamics of such
an extended object can be described, for example, in terms of the 
Polyakov--like action 
\beq
\label{pol}
I_n[g(\sigma),y(\sigma)]={T\over2}\int d^{n+1}\sigma \sqrt{g}\left(
g^{ij}\pa_i y^\mu \pa_j y^\nu G_{\mu\nu}(y) +(1-n)\right).
\eeq
The worldvolume metric $g_{ij}$ is treated as an independent variable
and $g\equiv -{\rm det}g_{ij}$. Its equation of motion, for $n\neq 1$,
leads to the induced metric
\beq\label{ind}
g_{ij}=\pa_iy^\mu \pa_jy^\nu G_{\mu\nu}(y).
\eeq
Substituting it back in \eref{pol} leads to the (at the classical level
equivalent) Nambu--Goto--like action $I_n[y]=T\int d^{n+1}\sigma
\sqrt{g}$, where $g_{ij}$ is now the induced metric. 

The action which describes the interaction between the system of branes
and the gauge degrees of freedom can then be written as 
\beq
\label{int}
I[A,C,a]=I_0[A,C,a]+\sum_{r=1}^{N_2}I_n^r +\sum_{r=1}^{N_1}I_{D-n-4}^r,
\eeq
where each of the individual kinetic terms for branes and dual branes
is of the form \eref{pol}. The action for the asymmetric theory is 
given by 
\beq
\widetilde I= I+{1\over2}\eta\int C^1C^2.
\label{asym}
\eeq

To complete the description of the dynamics, and specify the action
completely, one has to properly define the $C^I$--forms in terms of 
the dynamical variables of the branes. To provide this link it is
convenient to rely on the concept of Poincar\`e--duality, generalized
to the space of differential $p$--forms with distribution--valued components 
i.e. $p$--currents. In this space Poincar\`e--duality associates
to every $p$--dimensional hypersurface $\Sigma_p$ a $(D-p)$--form 
$\Phi_{\Sigma_p}$, such that 
\beq
\int_{\Sigma_p}\Psi_p=\int_{R^D} \Phi_{\Sigma_p} \Psi_p,
\label{pd}
\eeq
for any  $p$--form $\Psi_p$. The explicit expression for 
$\Phi_{\Sigma_p}$ is given in the appendix.
The  Poincar\`e--duality--map 
respects, in particular, the exact--forms $\leftrightarrow$ 
boundary--hypersurfaces correspondence, i.e.
\bea
&&\Sigma_p \rightarrow \Phi_{\Sigma_p} \qquad {\rm implies} \\
&&\pa \Sigma_p \rightarrow d\Phi_{\Sigma_p},
\eea
where $\pa$ indicates the boundary operator \footnote{See also \cite{LM};
for a mathematically precise formulation, involving chains, see 
\cite{deRham}.}. A basic consequence of Poincar\`e--duality is that   
the integral of $\Phi_{\Sigma_p}$ over a generic $(D-p)$--dimensional 
surface $\Sigma_{D-p}$ is an integer, counting the 
intersections with sign of $\Sigma_p$ with $\Sigma_{D-p}$. This implies, in 
particular, that the integral over $R^D$ of a product of
two such forms is also an integer,
\beq
\label{integer}
\int_{R^D} \Phi_{\Sigma_p}\Phi_{\Sigma_{D-p}}= \int_{\Sigma_p}
\Phi_{\Sigma_{D-p}}=N,
\eeq
which counts the number of intersections with sign of $\Sigma_p$ 
with $\Sigma_{D-p}$. 
Linear combinations of such $p$--forms with integer coefficients are called
{\it integer} forms.
The property \eref{integer} holds also for generic integer forms: the integral 
of a product of two integer forms is an integer, whenever the integral
is well defined. 

The forms $C^I$, whose defining property is $dC^I=J^I$ (see \eref{defc}),
can be determined as follows. The worldvolume of say the $r$-th $n$--brane 
is a closed $(n+1)$--dimensional hypersurface $\Sigma^2_r$, 
parameterized by $y^\mu_r(\sigma)$. Its Poincar\`e--dual is a closed 
integer $(D-n-1)$--form; we call it $J^2_r$, $dJ^2_r=0$. The total
current $J^2$, appearing in \eref{eq2}, is then given by
$$
J^2=\sum_{r=1}^{N_2}e^2_r J^2_r.  
$$  
Since $\Sigma^2_r$ is a closed $(n+1)$--dimensional hypersurface, 
due to the Poincar\`e lemma for currents \cite{deRham},
there exists an $(n+2)$--dimensional hypersurface $\Omega_r^2$, a
generalized ``Dirac--string", such
that $\pa\Omega_r^2=\Sigma^2_r$. Consequently, the Poincar\`e--dual of 
$\Omega^2_r$ is a $(D-n-2)$--form, we call it $C^2_r$, satisfying
$J^2_r=dC_r^2$. Therefore,
$$
C^2=\sum_{r=1}^{N_2} e^2_r C^2_r.               
$$
Since the same analysis applies also to the dual $(D-n-4)$--branes,
we have
\beq
J^I=\sum_{r=1}^{N_I}e^I_r J^I_r, \qquad
C^I=\sum_{r=1}^{N_I}e^I_r C^I_r.    
\label{param}
\eeq
However, the hypersurfaces $\Omega_r^I$ are not uniquely determined since, 
in general, there exist infinitely many hypersurfaces whose boundaries are
$\Sigma^I_r$. But under a Dirac--string change $\Sigma^I_r\rightarrow 
\widehat\Sigma^I_r$, we have $\widehat\Sigma^I_r-\Sigma^I_r=\pa 
\Lambda^I_r$, for some
hypersurface $\Lambda_r^I$. Under such a string change the $C$--fields
vary as
\beq
\widehat C^I -C^I=dW^I, \quad W^I\equiv\sum_{r=1}^{N_I}e^I_r W^I_r,
\label{defh}
\eeq
where the $W^I_r$, $(p^I-1)$--forms, are Poincar\`e--dual to 
$\Lambda^I_r$. These are precisely the transformations which generate the
Dirac--anomaly, see \eref{change}.

For what follows it is important to realize that the forms 
$J_r^I,C^I_r$ and $W^I_r$ are all {\it integer} forms. 

With the above determination of the $C^I$--fields, the action 
for the interacting system given in \eref{int}
becomes, actually, a functional of the {\it strings}
$C^I$, and not only of the worldvolumes, parameterized by 
$(y^I)^\mu_r(\sigma)$. But, what needs to be string--independent at the 
{\it quantum} level is rather the exponential $exp(iI)$ then the action 
itself. 
By definition, the response of the action under a string change is 
measured by the Dirac--anomalies . For the symmetric and asymmetric theories 
respectively, they can be evaluated using \eref{diracs}, \eref{diraca},
\eref{param} and \eref{defh}:
\bea
A_D&=& {1\over 2}\eta(-)^{D+n+1}\sum_{r,s}e_r^1e_s^2\int
                       \left[W^1_rJ^2_s+J^1_r W^2_s\right]\\
\widetilde A_D &=& \eta(-)^{D+n+1}\sum_{r,s}e_r^1e_s^2\int
                       W^1_rJ^2_s.
\eea
Quantum consistency constrains these anomalies to be integer multiples of 
$2\pi$. Since the integrands are products of integer forms, the integrals
in the expressions above are (positive or negative) integer numbers, and we 
obtain as quantization conditions for the charges, respectively for
the symmetric and asymmetric theory
\bea
\label{strong}
{1\over2}e_r^1e_s^2 &=&2\pi n_{rs}\\
e_r^1e_s^2 &=&2\pi \widetilde n_{rs},
\label{weak}
\eea
for each $r$ and $s$. 

If the charges satisfy the stronger condition \eref{strong}, then the 
symmetric and asymmetric theories coincide, actually. In this case we have, 
indeed,
$$
\widetilde I-I={1\over2}\eta\int C^1C^2={1\over2}
\eta\sum_{r,s}e_r^1e_s^2\int C^1_rC^2_s=2\pi m,
$$
with $m$ integer (here we used the fact that the $C^I_r$ are integer 
forms, too). 

We can conclude that a necessary and sufficient condition
for the quantum consistency of a system of interacting branes and dual
branes is the (weaker) Dirac--condition \eref{weak}, and that the 
corresponding theory is described by the (asymmetric) action $\widetilde I$. 
The theory based on the action $I$, with the consistency condition 
\eref{strong}, is just a special case of the asymmetric theory. 

Hence, in the case of branes/dual branes there is only {\it one} quantum 
mechanically consistent theory -- as anticipated above -- whose dynamics is 
represented by the asymmetric action $\widetilde I$, with the consistency
condition \eref{weak}.

We anticipated also that there is only {\it one} asymmetric theory in that
the two actions $\widetilde I_\pm=I\pm {1\over 2}\eta\int C^1C^2$ are 
equivalent. We see now that the equivalence stems from the fact that, 
due to \eref{weak}, 
the difference $\widetilde I_+-\widetilde I_-=\eta\int C^1C^2$ is an 
integer multiple of $2\pi$.

Once the functional $S\equiv exp(i\widetilde I)$ has been established to be 
string--independent it becomes a functional of only the brane
worldvolumes. Therefore, an action principle based on $S$ leads 
to equations of motion which are automatically string--independent and
respect, moreover, the PST--symmetries. All this information restricts
notably the form of the equations of motion for the branes, and their 
derivation is a mere exercise \footnote{Some caution is required if one
wants to take into account also configurations for which the string 
associated to a brane intersects the worldvolume of a dual brane; for
these configurations the consistent regularization procedure worked out 
in \cite{LM} for dyons in four dimensions applies equally well also in
the present case.}. For the $n$--branes and $(D-n-4)$--branes one 
obtains respectively the following equations of motion:
\bea\nonumber
T_r^2\,\qua y_r^\mu&=&\,\eta(-)^{nD}\,e_r^2\,{\pa y_r^{\nu_1}\over\pa\sigma^0}
\cdots  {\pa y_r^{\nu_{n+1}}\over\pa\sigma^n}\,
(K^1)^\mu{}_{\nu_{n+1}\cdots\nu_1}(y_r)\\ 
T_r^1\, \qua y_r^\mu&=&\,\eta(-)^{D+n+1}\,e_r^1\,
{\pa y_r^{\nu_1}\over\pa\sigma^0}
\cdots  {\pa y_r^{\nu_{D-n-3}}\over\pa\sigma^{D-n-4}}\,
(K^2)^\mu{}_{\nu_{D-n-3}\cdots\nu_1}(y_r). 
\label{lorentz}
\eea
The completely covariant laplacians on the worldvolumes are defined by
$$
\qua y^\mu
= {1\over \sqrt{g}}\pa_i(\sqrt{g}g^{ij}\pa_j y^\mu)+g^{ij}\pa_i y^\alpha
\pa_j y^\beta \Gamma^\mu_{\alpha\beta},
$$
where the metric $g^{ij}$ is
the induced one \eref{ind}, and $\Gamma^\mu_{\alpha\beta}$ is the affine 
target--space connection, evaluated on the brane. 

These are the expected equations of motion.
The appearance of the invariant tensors $K^I$, instead of the $F^I$, 
is dictated by the PST--symmetries, as explained above. 

\section{$S$--duality}

A basic property of the PST--action \eref{PST} is its invariance under
a simultaneous $S$--duality transformation of the vector potentials 
$A^I$. 

To prove this statement we give first
an equivalent representation for the third line in 
\eref{PST}. For this purpose we introduce a couple of (dual) gauge potentials
$\Lambda^I$, with degrees $D-p^I-1$, and a couple of $p^I$--forms $G^I$. 
With these new ingredients we write the equivalent action
\beq\label{equiv}
I_0[G,\Lambda,C,a]\equiv 
{1\over2}\int (G+C)\,{\cal P}(v)\,(G+C)+
\eta\left(C^1G^2-G^1C^2\right)
+\eta\left(d\Lambda^1G^2-G^1d\Lambda^2\right). 
\eeq
The equations of motion for $\Lambda^I$ give $dG^I=0$ $\rightarrow$
$G^I=dA^I$. Substituting this expression for $G^I$ in \eref{equiv}
one is back to the PST--action.

The $S$--transformed action, $\widehat I_0$, is defined by the 
functional integral over $G^I$,
\beq
e^{i\widehat I_0[\Lambda,C,a]}\equiv \int \{{\cal D}G\}\,
e^{iI_0[G,\Lambda,C,a]}.
\eeq
To evaluate the integral it is convenient to perform the shift 
$G^I\rightarrow G^I-C^I$,
\beq
\label{inter}
e^{i\widehat I_0[\Lambda,C,a]}= e^{i{\eta\over2}\int C^1d\Lambda^2
-d\Lambda^1 C^2}
 \int \{{\cal D}G\}\,
e^{{i\over2}\int G\,{\cal P}(v)\,G +\eta( (d\Lambda^1 +C^1)G^2
-G^1(d\Lambda^2 +C^2))}.
\eeq
The prefactor of the functional integral reproduces already the last term
in the third line of \eref{PST} (with the substitution $A^I\rightarrow
\Lambda^I$), and the (gaussian) functional integral itself depends only on the 
combinations $d\Lambda^I+C^I$. To perform it explicitly one has only to 
know the inverse of the operator ${\cal P}(v)$; apart from an overall 
factor of 4 and of some sign changes the structure of this operator is 
the same as the one of ${\cal P}(v)$. It is given by
\bea
\nonumber
{1\over 4}\,{\cal P}^{-1}(v)&=&
\left(\matrix{\eta(-)^n& 0\cr 
                       0&(-1)^{nD+n+1}\cr}\right)vi_v*
+\left(\matrix{0&\eta(-)^{nD}\cr 
    \eta(-)^{D+n+1}&0\cr}\right)vi_v\\
&\phantom{=}&\nonumber\\
 &\phantom{=}&  +{1\over2}
  \left(\matrix{0&\eta(-)^{nD+n+1}\cr 
\eta&0\cr}\right).
\eea
This leads to the simple result
$$
\int \{{\cal D}G\}\,
e^{{i\over2}\int G\,{\cal P}(v)\,G +\eta( (d\Lambda^1 +C^1)G^2
-G^1(d\Lambda^2 +C^2))}=
e^{{i\over2}\int (d\Lambda+C){\cal P}(v)(d\Lambda+C)}.
$$
Using this in \eref{inter} we arrive at
$$
\widehat I_0[\Lambda,C,a]=I_0[\Lambda,C,a],
$$
i.e. again the PST--action for the couple $\Lambda^I$.

\section{Particular cases}

We discuss in this section some particular relevant cases: chiral bosons,
dyons in twice even dimensions and dyons in twice odd dimensions. The 
actions which govern the dynamics of these objects are obtained by 
specializing the general action \eref{PST} to these cases.

\subsection{Chiral bosons and self--dual branes in $D=4K+2$}

We begin by considering gauge potentials with (anti)self--dual 
field strengths -- (anti)chiral bosons -- which exist in dimensions $D=4K+2$. 
In this case we have a single $2K$--form gauge potential $A$ and a single
$(2K+1)$--form  string $C$. The curvature is also a  $(2K+1)$--form, 
$F=dA+C$, and satisfies as
equations of motion the (anti)self--duality relations
\beq
F=\pm *F,
\label{chiral}
\eeq
for chiral and antichiral bosons respectively. The consistency of these
equations is guaranteed by the fact that now
$$
*^2=\eta=+1.
$$

The branes coupled to these vector potentials are $(2K-1)$--branes 
$(n=2K-1=\,odd)$, and coincide with the dual branes. Correspondingly we 
have now only a single type of (electric = magnetic) charges 
$$
e_r^1=e_r^2\equiv e_r.
$$
These branes can couple to chiral or antichiral bosons.
The actions for these systems can be obtained by enforcing in \eref{PST} the 
identifications
\bea\nonumber
A^1=A, &\quad& A^2=\pm A\\
C^1=C, &\quad& C^2=\pm C,
\label{identif}
\eea
which lead to $F^1=F$, $F^2=\pm F$. With these substitutions one
obtains, for chiral and antichiral bosons respectively, the 
actions\footnote{It is understood that here and in the cases treated below
eventually one has to add also the kinetic terms for the branes
$\sum_r T_r\int d^{n+1}\sigma\sqrt{g_r}$.} 
\beq
\label{azch}
I_0^\pm[A,C,a]=\pm \int Fvi_v(F\mp *F) +CdA.
\eeq
It is important, but easy, to realize that
the above identifications are compatible with the PST--symmetries. 
Therefore, $a$ is again an auxiliary field and the equation of motion
for $A$ is $d(vf^\pm)=0$, where $f^\pm\equiv i_v(F\mp *F)$. With the
usual gauge fixing of the symmetry \eref{2} for the single potential $A$,
it reduces just to \eref{chiral}.

Since in the present case $C$ is an odd form the term $\int CC$ is 
identically zero and we have only a single type of theories. The 
Dirac--anomalies can be obtained inserting \eref{identif} in \eref{diracs}
(or \eref{diraca}) with the result
\beq
\label{diracch}
A_D^\pm=\Delta I_0^\pm=\pm\int WJ.
\eeq
Clearly they can be obtained directly also from \eref{azch}. The
variations are here $\Delta C=dW$, $\Delta A=-W$, and $J\equiv dC$.

If we  have a number $N$ of dynamical branes with charges $e_r^+$
($e_r^-$) coupled to chiral (antichiral) bosons,   
the expressions for $J$, $C$ and $W$ are, as in the previous section,
$J=\sum_r e_r^\pm J_r$, $C=\sum_r e_r^\pm C_r$, $W=\sum_r e_r^\pm W_r$, 
with $J_r$, $C_r$ and $J_r$ integer forms. The Dirac--anomalies become then 
$$
A_D^\pm=\pm  \sum_{r,s} e_r^\pm e_s^\pm \int W_rJ_s,
$$
and the quantization condition is of the Dirac--type
\beq
\label{selfdual}
e_r^\pm e_s^\pm =2\pi n_{rs},
\eeq
for each $r$ and $s$. 
A class of solutions \footnote{Notice, however, that \eref{sol} does not 
correspond to 
the most general solution of \eref{selfdual}.} of these conditions is given by
\beq\label{sol}
e_r^\pm=\sqrt{2\pi L}\cdot n_r,
\eeq
where $L$ is a fixed integer and the $n_r$ are integers, too.

The effective actions amount now to
\beq
\Gamma_\pm[C]= \int\left(-J\, {*\over \qua}\, J\pm J\,{\delta\over 
\qua}\,C\right)=\sum_{r,s}e_r^\pm e_s^\pm\int\left(
-J_r\, {*\over \qua}\, J_s\pm J_r\,{\delta\over 
\qua}\,C_s\right).
\label{chirali}
\eeq
The first term represents the standard Coulomb interactions between
the branes; the second
term represents the mixed interactions which we called previously 
electric--magnetic interactions. The striking feature in this case is that
both types of interactions are weighted by the same effective coupling 
constant $e_re_s$ which is, actually, a mixed coupling constant. For branes
coupled to chiral bosons there is indeed no way to distinguish between
magnetic and electric couplings. This is clearly due to the fact that 
a brane which couples electrically to a chiral boson through $d*F=J$, 
carries necessarily also a  magnetic coupling, because $F=*F$ implies
$dF=J$. Therefore an electric brane is automatically turned in a magnetic
brane and viceversa. In the case of dyons, considered in the next 
section, the mutual interactions will exhibit a rather different behaviour.

Another peculiar feature of the above expression for the effective action
is represented by the self--interaction of the $r$--th brane, corresponding
to the terms with $r=s$. In the terms describing the Coulomb 
self--interactions, $J_r{*\over\qua}J_r$, it leads to the 
standard short distance ultraviolet divergences; but also the mixed 
self--interactions, $\pm J_r{\delta\over\qua}C_r$,
show up ultraviolet divergences. Whereas the formers, being
strictly independent on the Dirac--strings, can 
be regularized in a standard manner, the regularization of the latters has
to be handled carefully because it has to be chosen such that the
Dirac--anomaly remains an integer multiple of $2\pi$. We will come back 
to this point in the following subsection.

\subsection{Dyons in $D=4K$}

Dyonic branes are branes which carry electric {\it and} magnetic charge. They
can live only in an even dimensional space--time, and in $D=4K$ they
become $(2K-2)$--branes, $(n=2K-2=\,even)$. A system of $N$ such branes
is characterized by the charge vectors $e_r^I=(e_r^1,e_r^2)$, 
$r=1,\cdots,N$, and by the total currents 
\beq
\label{curr}
J^I=\sum_{r=1}^Ne^I_r J_r,
\eeq
where the $J_r$ are the Poincar\`e--duals of the brane worldvolumes. 
The forms $C^I$ and $W^I$ are expressed as sums completely analogous to 
\eref{curr}. Notice 
that also in this case there is no distinction between branes and dual 
branes, since they are all $(2K-2)$--branes and carry electric as well
as magnetic charge. 

The action for such a system is given by \eref{PST}, in terms of two
vector potentials $A^I=(A^1,A^2)$, and we have only to specify the signs. 
In $D=4K$ we have, in particular,  
$$
*^2=\eta=-1.
$$

The unique difference to the case presented in sections three and four 
relies in the different parametrizations of the currents, respectively 
\eref{param} and \eref{curr}, and the corresponding differences in the 
parametrizations of $C^I$ and $W^I$. We treat now the symmetric and 
asymmetric theories separately.

{\it Symmetric theory.} 
Keeping the notation of the third line of \eref{PST} we obtain for the
action
\beq\label{azeven}
I_0[A,C,a]=
{1\over2}\int\left[ F\,{\cal P}(v)\,F +dA
\left(\matrix{0&1\cr -1&0\cr}\right)C\right],
\eeq
where now
\beq
{\cal P}(v)=
\left(\matrix{1& 0\cr 0&1\cr}\right)vi_v*
+\left(\matrix{0&1\cr -1&0\cr}\right)\left(vi_v-{1\over2}\right).
\eeq
Since $A^1$ and $A^2$ are now forms of the same degree, $A^I$ can be 
considered as an  $SO(2)$--doublet. If we consider also the charge 
vectors $e^I_r$ as $SO(2)$--doublets, the action $I_0$ is 
manifestly invariant under the continuous duality group $SO(2)$. This is 
due to the fact that the $2\times2$ matrices appearing in \eref{azeven},
the identity and the antisymmetric tensor, are $SO(2)$--invariant.

Let us now see what happens to the Dirac--anomaly, which triggers
the quantum consistency of the system of dyons. Eq. \eref{diracs} gives
\beq
A_D={1\over2}\int W \left(\matrix{0&1\cr-1&0\cr}\right)J
    ={1\over2}\sum_{r,s}(e_r^1e_s^2-e_r^2e_s^1)\int W_rJ_s.
\label{diracnew}
\eeq
The sign flip in $A_D$, with respect to \eref{diracs}, is due to the fact
that $J^I$ and $W^I$ are forms of {\it odd} degree. Since the integrals 
$\int W_rJ_s$ are integers we obtain the quantization conditions
\beq
\label{ds}
{1\over2}(e_r^1e_s^2-e_r^2e_s^1)=2\pi n_{rs},
\eeq
which are the known Dirac--Schwinger quantization conditions for the 
charges of dyons in $D=4K$. The effective action becomes, from 
\eref{gamma} 
\bea
\nonumber
\Gamma[C]&=&{1\over2}\int\left[J {*\over \qua}\, 
\left(\matrix{1&0\cr 0& 1\cr}\right)\,J 
+ J {\delta\over\qua}\, \left(\matrix {0&-1\cr 1&0\cr}\right)\,C\right]\\
&=&{1\over 2}\sum_{r,s} \int
\left[(e_r^1e_s^1+e_r^2e_s^2)J_r\,{*\over\qua}\,J_s+(e_r^2e_s^1-e_r^1e_s^2)
J_r\,{\delta\over\qua}\,C_s\right].
\label{gamma1}
\eea
The effective coupling constant matrix $\gamma_{rs}=e_r^1e_s^1+e_r^2e_s^2$
for the Coulomb interactions differs now from the one of the mixed
interactions, $\beta_{rs}=e_r^2e_s^1-e_r^1e_s^2$, but both matrices 
are $SO(2)$--invariant. 
Notice in particular the absence of mixed self--interactions; this is due 
to the antisymmetry of the mixed coupling constant matrix $\beta_{rs}$.
The presence of such interactions will represent the major difference 
between the symmetric and asymmetric theory. 

The generalized Lorentz--force laws can be deduced from $exp(iI)$. Here
$I$ is obtained from $I_0$ adding the kinetic terms for the dyons, as in 
\eref{int}, but now with a single sum. Instead of \eref{lorentz} one obtains 
now\footnote{Formally one has to sum the right hand sides of 
\eref{lorentz}}
\beq
\label{lorentzd}
T_r\qua y_r^\mu= {\pa y_r^{\nu_1}\over\pa\sigma^0}
\cdots  {\pa y_r^{\nu_{n+1}}\over\pa\sigma^n}\,
\left[e_r^1 K^2- e_r^2 K^1\right]^\mu{}_{\nu_{n+1}\cdots\nu_1}(y_r).
\eeq
The square bracket reflects again the $SO(2)$--invariance of the symmetric
theory.

{\it Asymmetric theory}. The action for this theory is given by 
\beq
\widetilde I_0=I_0-{1\over2}\int C^1C^2=
{1\over2}\int\left[ F\,{\cal P}(v)\,F +dA
\left(\matrix{0&1\cr -1&0\cr}\right)C-C^1C^2\right].
\eeq
The Dirac--anomaly can be read from \eref{diraca}
\beq
\label{diracold}
\widetilde A_D=\int W^1J^2=\sum_{r,s}e_r^1e_s^2\int W_rJ_s.
\eeq
It becomes an integer multiple of $2\pi$ if
\beq
\label{d}       e_r^1e_s^2=2\pi \widetilde n_{rs},
\eeq
for each $r$ and $s$. This is Dirac's original quantization condition
and it has to be compared with the Dirac--Schwinger condition \eref{ds} of the
symmetric theory. None of the two conditions implies the other. The most 
relevant difference is that Dirac's original condition requires 
quantization also for the electric and magnetic charges of a single dyon:
$e_r^1e_r^2=2\pi \widetilde n_{rr}$, which allows for an electric--magnetic
self--interaction of the $r$--th dyon, as we will see in a moment. 
Moreover, the Dirac--Schwinger conditions admit solutions for the charges 
with $\vartheta$--angles while Dirac's original condition does not
allow for such angles and, to introduce them in the asymmetric theory,
one has to amend the action $\widetilde I_0$ with a 
$\vartheta$--term, see section seven.

The principal reason for these differences is that the asymmetric theory 
looses the  $SO(2)$--duality invariance of the symmetric
theory, due to the presence of the additional term $-{1\over2}\int C^1C^2$
in the action. 
The unbroken duality sub--group of $SO(2)$ which survives is the discrete 
group $Z_4$, generated by
\beq
\label{z4}
{e_r^1\choose e_r^2} \rightarrow \left(\matrix{0&1\cr-1&0\cr}\right)
{e_r^1\choose e_r^2}.
\eeq
Since under this transformation we have $C^1\rightarrow C^2$, 
$C^2\rightarrow -C^1$, the action changes as\footnote{$I_0$ is invariant
and $C^1C^2$ changes its sign.}
\beq
\label{tras} 
\widetilde I_0\rightarrow \widetilde I_0 +\int C^1C^2.
\eeq
Therefore, strictly speaking $\widetilde I_0$ is invariant only under the 
(trivial) duality group $Z_2$, generated by minus the identity matrix. 
However, at the quantum level it is sufficient that the exponentiated 
action is invariant. Indeed, due to \eref{d} and to the fact that the 
$C_r$ are integer forms, the integral 
$\int C^1C^2 =\sum_{r,s}e_r^1e_s^1\int C_rC_s$  is an integer
multiple of $2\pi$, 

The integral $\int C_rC_r$ is, actually, ill--defined 
because it counts the number of intersections of two ``coinciding"
surfaces, which is infinite. One can regularize this term by displacing
the position of one surface by an arbitrary infinitesimal vector 
$\varepsilon^\mu$ orthogonal to the surface. This leads to a framed
$(n+2)$--form $C_r^\varepsilon$, and the integral $\int C_rC_r^\varepsilon$
is now well--defined and integer. 

Eventually the difference between the two theories is most clearly exhibited
by the effective actions. For the asymmetric theory we obtain from 
\eref{gammaas}
\bea\nonumber
\widetilde\Gamma[C]&=&\int\left[{1\over2}J {*\over \qua}\, 
\left(\matrix  {1&0\cr 0& 1\cr}\right)\,J 
+ J^2 {\delta\over\qua}\, C^1\right]=
\sum_{r,s} \int
\left[{1\over 2}(e_r^1e_s^1+e_r^2e_s^2)J_r\,{*\over\qua}\,J_s+e_r^2e_s^1
J_r\,{\delta\over\qua}\,C_s\right]\\
&=&
{1\over 2}\sum_{r,s} \int
\left[(e_r^1e_s^1+e_r^2e_s^2)J_r\,{*\over\qua}\,J_s+(e_r^2e_s^1-e_r^1e_s^2)
J_r\,{\delta\over\qua}\,C_s\right] \nonumber\\
&&+\sum_r e_r^1e_r^2\int J_r\,{\delta\over \qua}\,C_r 
-{1\over 2}\sum_{r>s} (e_r^1e_s^2+e_r^2e_s^1)\int C_rC_s.
\label{gamma2}
\eea
To obtain the last expression we used the Hodge decomposition of the
D'Alambertian in the form
\beq
\label{ids}
\int\left[J_r\,{\delta\over\qua}\,C_s +J_s\,{\delta\over\qua}\,C_r\right]
=-\int C_rC_s,
\eeq
which holds for arbitrary even forms $C_{r,s}$ in $D=4K$ with
$J_{r,s}=dC_{r,s}$, thus separating the electric--magnetic interactions
in antisymmetric, diagonal and symmetric contributions (w.r.t. $r$ 
and $s$).
This expression has to be compared with the effective action
for the symmetric theory \eref{gamma1}. The second line in \eref{gamma2}
equals the effective action for the symmetric theory, the third line
represents two additional terms. The second term is made out of 
integer multiples of $\pi$, due to \eref{d}, and gives in general non
vanishing contributions to the (exponentiated) effective action. The first
term describes the above mentioned electric--magnetic self--interaction of 
the $r$--th dyon, and it needs a regularization. If we use a framing 
regularization as above, we obtain
\beq
\label{reg}
\Gamma_{self}
\equiv\sum_r e_r^1e_r^2\int J_r\,{\delta\over \qua}\,C_r^\varepsilon
={1\over2}\sum_r e_r^1e_r^2\int \left[J_r\,{\delta\over\qua}\,C_r^\varepsilon 
-J_r^\varepsilon\,{\delta\over\qua}\,C_r -C_rC_r^\varepsilon \right],
\eeq
where $J_r^\varepsilon =dC_r^\varepsilon$. Here we used again \eref{ids};
notice, however, that in absence of a regularization this formula would 
have led to the meaningless result 
$\int J_r\,{\delta\over \qua}\,C_r=-{1\over 2}\int C_rC_r$. The first two
terms of $\Gamma_{self}$, which seem to cancel each other as 
$\varepsilon\rightarrow 0$,
converge actually to a finite (non integer) non vanishing result. The third
term, on the other hand, is an integer multiple of $\pi$ and gives a non 
vanishing contribution to the (exponentiated) effective action, too. The
important feature of this regularization is that it keeps the 
Dirac--anomaly an integer multiple of $2\pi$. Under 
$\Delta C_r^\varepsilon= dW_r^\varepsilon$, using the definition
of $\Gamma_{self}$  and $\qua=-(d\delta+\delta d)$, we have 
$$
\Delta\Gamma_{self}=\sum_r e_r^1e_r^2\int J_rW_r^\varepsilon,
$$
which is again a well defined integer multiple of $2\pi$, due to \eref{d}
for $r=s$.

The major physical effect of this self--interaction term for
dyons in $D=4$ in the asymmetric theory
is the spin--statistics transmutation of the
$r$--th dyon from a boson to 
a fermion, and viceversa, if the integer $\widetilde n_{rr}$ appearing
in the quantization condition of the asymmetric theory is {\it odd}
\cite{LM2}. This transmutation does not occur in the symmetric theory.

It is instructive to analyse the implementation of the $Z_4$--invariance 
of the asymmetric theory a the level of the (regularized) effective action. 
According to its generator \eref{z4} we have to perform the replacements
$e_r^1\rightarrow e_r^2$, $e_r^2\rightarrow -e_r^1$ and also 
$C_r^\varepsilon\leftrightarrow C_r$, which implies  
$J_r^\varepsilon\leftrightarrow J_r$. The second line in \eref{gamma2}
is manifestly invariant under $SO(2)$ and hence also under $Z_4$. The
second term of the third line changes its sign, but, being an integer
multiple of $\pi$, it changes by an integer multiple of $2\pi$. The
first term of the second line has to be substituted by its regularized 
version \eref{reg}. Since $e_r^1e_r^2\rightarrow -e_r^1e_r^2$ the first
two terms in the square bracket compensate this sign change, due to 
$C_r^\varepsilon\leftrightarrow C_r$, and the third term corresponds to 
an integer multiple of $\pi$ and changes by an integer multiple of $2\pi$,
as above.

We observe finally that if the quantization conditions  \eref{ds} and
\eref{d} hold simultaneously, then the difference between the symmetric and
asymmetric theories reduces just to $\Gamma_{self}$, a part from an 
integer multiple of $2\pi$. This is due to the trivial identity
\beq
\label{trivial}
{1\over 2}(e_r^1e_s^2+e_r^2e_s^1)={1\over 2}(e_r^1e_s^2-e_r^2e_s^1)
+ e_r^2e_s^1= 2\pi N_{rs}, 
\eeq
where we used both quantization conditions.

The generalized Lorentz--force law for the asymmetric
theory is identical to the one of the symmetric theory, \eref{lorentzd}, 
and it is in particular $SO(2)$--invariant. We would like to stress that
the distinction between symmetric and
asymmetric theories arises really at the quantum level, at the 
classical level they are identical. 

The analysis of the distinctive features of the symmetric and asymmetric 
theories performed above, in particular the correct identification of the
self--interaction in the asymmetric theory, refines the corresponding 
analysis performed in \cite{LM}. 

\subsection{Dyons in $D=4K+2$}

This case can be treated in the same way as the preceding one. The branes
are $(2K-1)$--branes, $(n=2K-1=\,odd)$, and $*^2=\eta=+1$. The $F$'s and
$C$'s are now {\it odd} forms. 
 
With respect to the $D=4K$ case we have only some sign changes in due 
places. Nevertheless, the main feature of dyons in $D=4K+2$ is that the
symmetric and asymmetric theories are more similar then the corresponding
theories in $D=4K$. This is due to the fact that, as we will see, 
in $D=4K+2$ the two theories are invariant under the {\it same} duality 
group, i.e. $Z_2\times Z_2'$.

{\it Symmetric theory.} The action for the symmetric theory becomes
\beq\label{azodd}
I_0[A,C,a]=
{1\over2}\int\left[F\,{\cal P}(v)\,F -dA
\left(\matrix{0&1\cr 1&0\cr}\right)C\right],
\eeq
where 
\beq
{\cal P}(v)=
-\left(\matrix{1& 0\cr 0&1\cr}\right)vi_v*
+\left(\matrix{0&1\cr 1&0\cr}\right)\left(vi_v+{1\over2}\right).
\eeq
The sign flip in the last term of \eref{azodd}, with respect to \eref{PST},
is due to the fact that $dA^I$ and $C^I$ are here odd forms.

The $2\times 2$ matrices involved in \eref{azodd} are the identity 
and the matrix $\left(\matrix{0&1\cr 1&0\cr}\right)$. These matrices
are left invariant by the discrete duality group $Z_2\times Z_2'$, generated
by 
$$
g=-\left(\matrix{1&0\cr0&1\cr}\right)\quad {\rm and}\quad
g'=\left(\matrix{0&1\cr1&0\cr}\right),
$$
respectively.
Therefore, the symmetric theory admits the duality--symmetry group 
$Z_2\times Z_2'$, the first factor being almost trivial \cite{DESER}.

The Dirac--anomaly is 
\beq
A_D={1\over2}\int W \left(\matrix{0&1\cr 1&0\cr}\right)J
    ={1\over2}\sum_{r,s}(e_r^1e_s^2+e_r^2e_s^1)\int W_rJ_s,
\eeq
and leads to the quantization condition
\beq
\label{ds2}
{1\over2}(e_r^1e_s^2+e_r^2e_s^1)=2\pi n_{rs},
\eeq
first noted in \cite{DESER}, which shows the relative plus sign instead 
of the minus sign of dyons in $D=4K$. In particular, we have now
also a quantization condition between the charges $e_r^1$ and $e_r^2$ of 
the $r$--th dyonic brane, $e_r^1e_r^2=2\pi n_{rr}$, allowing for a 
self--interaction. The effective action is, indeed 
\bea
\nonumber
\Gamma[C]&=&-{1\over2}\int J {*\over \qua}\, 
\left(\matrix{1&0\cr 0& 1\cr}\right)\,J 
-J {\delta\over\qua}\, \left(\matrix {0&1\cr 1&0\cr}\right)\,C\\
&=&
-{1\over 2}\sum_{r,s} \int
\left[(e_r^1e_s^1+e_r^2e_s^2)J_r\,{*\over\qua}\,J_s-(e_r^2e_s^1+e_r^1e_s^2)
J_r\,{\delta\over\qua}\,C_s\right]\nonumber\\
&=&
-{1\over 2}\sum_{r,s} \int
\left[(e_r^1e_s^1+e_r^2e_s^2)J_r\,{*\over\qua}\,J_s\right]
+{1\over2}\sum_{r\neq s}\int\left[(e_r^2e_s^1+e_r^1e_s^2)
J_r\,{\delta\over\qua}\,C_s\right]+\Gamma_{self},\nonumber\\
&&
\label{gamma3}
\eea
where
\beq
\label{self1}
\Gamma_{self}=\sum_r \int e_r^1e_r^2\,J_r\,{\delta\over\qua}\,C_r^\varepsilon.
\eeq
We have again standard Coulomb interactions\footnote{
The minus sign of the Coulomb interactions is due to our conventions
regarding the Hodge--dual.} and mixed interactions weighted
by $Z_2\times Z_2'$--invariant coupling constant matrices.
But the main difference
w.r.t. the symmetric theory in $D=4K$ is represented by the presence of 
the electric--magnetic self--interactions
of the dyonic branes, represented by $\Gamma_{self}$, which we have 
regularized as above to preserve the Dirac--anomaly. 
In $D=4K+2$ the identity \eref{ids} is replaced by 
\beq
\label{ida}
\int\left[J_r\,{\delta\over\qua}\,C_s -J_s\,{\delta\over\qua}\,C_r\right]
=-\int C_rC_s,
\eeq
because the $C_r$'s are here {\it odd} forms. This allows to decompose 
the self--interaction as
\beq
\Gamma_{self}=
{1\over2}\sum_r e_r^1e_r^2\int \left[J_r\,{\delta\over\qua}\,C_r^\varepsilon 
+J_r^\varepsilon\,{\delta\over\qua}\,C_r -C_rC_r^\varepsilon \right].
\eeq
This time it is the last term which naively would go to zero as 
$\varepsilon\rightarrow 0$, because the $C_r$ are odd forms; but due
to the framing it converges to an integer multiple of $\pi$ which in 
general is non vanishing. 

For what concerns the duality--symmetries, under $Z_2$ the effective action 
is trivially invariant, while under the generator $g'$ of $Z_2'$ we have
$e_r^1 \leftrightarrow e_r^2$ and $C_r\leftrightarrow 
C_r^\varepsilon$. The unique term of $\Gamma$ which transforms non
trivially is the last term of $\Gamma_{self}$ which changes its sign; but 
since it is an integer multiple of $\pi$ $exp(i\Gamma)$ is invariant.

The generalized Lorentz--force equations are 
\beq
T_r\,\qua y_r^\mu= {\pa y_r^{\nu_1}\over\pa\sigma^0}
\cdots  {\pa y_r^{\nu_{n+1}}\over\pa\sigma^n}\,
\left[e_r^1 K^2+e_r^2 K^1\right]^\mu{}_{\nu_{n+1}\cdots\nu_1}(y_r),
\eeq
and they are $Z_2\times Z_2'$--invariant.
 
{\it Asymmetric theory}. The action is 
\beq
\widetilde I_0=I_0+{1\over2}\int C^1C^2,
\eeq
with Dirac--anomaly
\beq
\widetilde A_D=\int W^1J^2=\sum_{r,s}e_r^1e_s^2\int W_rJ_s.
\eeq
It leads to the quantization conditions
\beq
\label{d2}
e_r^1e_s^2=2\pi \widetilde n_{rs},
\eeq
which are Dirac's original ones, and they allow for self--interactions, too.
The asymmetric action is trivially invariant under $Z_2$, 
while under $g'$ it transforms as
\beq
\label{trasodd} 
\widetilde I_0\rightarrow \widetilde I_0 -\int C^1C^2,
\eeq
and $exp(i\widetilde I_0)$ is invariant due to \eref{d2}.
The effective action is 
\bea
\nonumber
\widetilde\Gamma[C]&=&-\int\left[ {1\over2}J {*\over \qua}\, 
\left(\matrix  {1&0\cr 0& 1\cr}\right)\,J 
-J^2 {\delta\over\qua}\, C^1\right]\\
&=& \sum_{r,s} \int
\left[-{1\over 2}(e_r^1e_s^1+e_r^2e_s^2)\,J_r\,{*\over\qua}\,J_s 
+e_r^2e_s^1\,
J_r\,{\delta\over\qua}\,C_s\right]\nonumber\\
&=& \Gamma[C]+{1\over 2}\sum_{r>s}\, (e_r^1e_s^2-e_r^2e_s^1)\int C_rC_s.
\label{gamma4}
\eea
Here $\Gamma[C]$ denotes the effective action of the symmetric theory
\eref{gamma3}, and we used again \eref{ida}. In particular, the 
self--interaction is present also here, and the unique difference 
between $\Gamma$ and $\widetilde\Gamma$ is represented by the last sum
in \eref{gamma4}, which amounts to an integer multiple of $\pi$. Therefore,
also $exp(i\widetilde\Gamma)$ is invariant under $Z_2\times Z_2'$. 
The generalized Lorentz--force law is the same as for the symmetric theory.

If both quantization conditions \eref{ds2} and \eref{d2} hold 
simultaneously, then an argument analogous to \eref{trivial} shows that
$exp(i\Gamma)=exp(i\widetilde\Gamma)$, and the two theories become
quantum mechanically indistinguishable.
 
We noted above that in $D=4$, for the asymmetric theory, the 
self--interaction leads to a spin--statistics transmutation; the
question whether a similar effect occurs also in higher even dimensions, 
e.g. $D=6$, is still under investigation. What we have shown here is that
{\it if} it occurs in $D=4K+2$, then the effect is present in both type of 
theories. 

Concluding we can say that the dynamics of dyons in an even dimensional 
space--time can be described at the quantum level by two inequivalent
theories. The asymmetric theory requires as consistency condition
a Dirac--type quantization condition and the symmetric one a 
Dirac--Schwinger--type quantization condition. At the quantum 
level for $D=4K$ the symmetric
theory is $SO(2)$--invariant and the asymmetric one $Z_4$--invariant, while
for $D=4K+2$ both types of theories are invariant under $Z_2\times Z_2'$.
 
\subsubsection{Chiral factorization}

A characteristic property of $(4K+2)$--dimensional space--times 
is the existence of chiral and antichiral bosons. This allows to 
separate a $2K$--form gauge potential in its chiral and 
antichiral part. Chiral and antichiral bosons should couple
respectively to the ``chiral" dyonic charges
$$
e_r^\pm\equiv{1\over2}(e_r^1\pm e_r^2).
$$

We want here to show how this factorization occurs naturally 
in the covariant action $I_0$  of the symmetric theory, 
given in \eref{azodd}. It suffices to define 
$$
A^\pm={1\over2}(A^1\pm A^2), \quad C^\pm={1\over2}(C^1\pm C^2).
$$
A short calculation gives then
\bea
I_0[A,C,a]&=&
{1\over2}\int\left[F\,{\cal P}(v)\,F -dA
\left(\matrix{0&1\cr 1&0\cr}\right)C\right]\\
&=&I^+_0[A^+,C^+,a] +I^-_0[A^-,C^-,a],
\eea
where the actions for chiral bosons $I_0^\pm$ are given in \eref{azch}.
The effective action factorizes similarly:
$$
\Gamma[C]=\Gamma_+[C]+\Gamma_-[C],
$$
where $\Gamma_\pm [C]$ is given in \eref{chirali}.
 
The action for the asymmetric theory admits only a partial factorization
and contains also a mixed term
$$
\widetilde I_0[A,C,a]=I^+_0[A^+,C^+,a] +I^-_0[A^-,C^-,a] +\int C^-C^+.
$$
Correspondingly one has $\widetilde \Gamma[C]= 
\Gamma_+[C]+\Gamma_-[C]+\int C^-C^+$. There remains, however, a link
between chiral and antichiral currents since the quantization conditions
\eref{ds2} and \eref{d2}, in terms of the chiral charges, read
\bea
e^+_re^+_s-e^-_re^-_s&=&2\pi n_{rs}\quad {\rm (symmetric)}\\
e^+_re^+_s-e^-_re^-_s + e^-_re^+_s-e^+_re^-_s&=&2\pi 
\widetilde n_{rs} \quad {\rm (asymmetric)},
\eea
and mix up chiral with antichiral charges. These conditions are less 
restrictive then the ones for chiral branes, i.e. 
$e_r^\pm e_s^\pm=2\pi n_{rs}$, and admit therefore more solutions.
The case of chiral branes corresponds to $e_r^-={1\over 2}
(e^1_r-e^2_r)=0=C^-$, $e_r^+=e_r^1$, meaning that the electric charges 
equal the magnetic ones. In this case the antichiral boson $A^-$ decouples, 
it becomes a free field, and one recovers the quantization conditions 
\eref{selfdual} for chiral (self--dual) branes.

The invariant field strengths corresponding to the chiral and antichiral 
bosons are given by $K^\pm={1\over2}(K^1\pm K^2)$ and satisfy 
$$
*K^\pm=\pm K^\pm.
$$

\section{More general couplings}

In this section we present generalizations of the PST--action which 
describe couplings to additional fields, and address the issue of 
$\vartheta$--angles. 

We recall
that for all models discussed in this paper, except for chiral
bosons, equivalent formulations in terms of a single gauge potential
are obtained by dropping in \eref{PST} in the first
line the term $f^1*f^1$, or in the second line the term $f^2*f^2$.
In the first case one obtains a formulation in terms of the potential
$A^1$, and in the second case a formulation in terms of the (dual) 
potential $A^2$. The principal drawback of formulations with a single 
gauge potential is that duality symmetries are not manifest; they can be 
realized only through non--local transformations \cite{LM}. 

Formally these formulations can be obtained from the PST--formulation 
in the following way. First one has to $S$--dualize the scalar auxiliary 
field $a$ to an auxiliary $(D-2)$--form $a_{D-2}$, obtaining thus the
so called ``dual" PST--action \cite{MPS}, again in terms of the two potentials
$A^1$ and $A^2$, $I_0^{dual}[A^I,C,a_{D-2}]$. Then one has to perform
the functional integration over $A^2$ (or $A^1$) of $exp(iI_0^{dual})$.
After this functional integration the field 
$a_{D-2}$ decouples, and one obtains the Schwinger--like action in terms 
of a single gauge potential (for more details see \cite{LM}). 

\vskip0.3truecm\noindent
{\it Gravitation.} The dynamics of the interacting system 
branes/potentials described in the preceding sections was assumed to
occur in a background with metric $G_{\mu\nu}(x)$. Implicitly all indices 
were raised and lowered with this metric. Diffeomorphism invariance 
is achieved thanks to the manifest Lorentz--invariance of the PST--method
in the flat case: the minimal--coupling--recipe gives therefore rise to 
manifestly diffeomorphism invariant
actions. This is one of the fundamental advantages of the PST--approach.
Moreover, the Dirac--anomaly is a topological invariant i.e. metric 
independent, and so the relevant quantization conditions ensure
Dirac--string independence of the exponentiated actions also in the 
presence of gravitation. To introduce a dynamical gravitational field
it is therefore sufficient to add the Einstein--Hilbert action.

\vskip0.3truecm\noindent
{\it Coupling to other fields.} The PST--action admits also consistent
couplings of the system gauge--potentials/branes to other fields. These
couplings are represented in general by a couple of $p^I$--forms $L^I$. 
$L^1(L^2)$ couples electrically to $A^2(A^1)$ and magnetically to 
$A^1(A^2)$. In supergravity theories, for example, the (composite)
fields $L^I$ 
correspond to Chern--Simons forms, made out of other $p$--form potentials or
of anomaly--cancelling terms, or to bilinears in the fermions. 
The modified dynamics is then expressed by the equations
\bea
H^I&\equiv& F^I+L^I=dA^I+C^I+L^I\nonumber\\
H^1&=&*H^2.
\label{proto}
\eea
Since we have now $dH^I=J^I+dL^I$, the forms $dL^I$ must be (gauge) 
invariant forms. Actually, the fields $L^I$ 
can depend also on the brane worldvolumes themselves; in this case 
the $L^I$ are required to be invariant under Dirac--string changes.

The action which takes the new couplings consistently into 
account is given for the symmetric theory by
\bea\nonumber
I_L&=&I_0[A,C+L,a]+\eta{1\over2}\int\left(L^1C^2-C^1L^2\right)\\
&=&{1\over2}\int\left[H\,{\cal P}(v)\,H +\eta
\left((C^1+L^1)dA^2-dA^1(C^2+L^2)+L^1C^2-C^1L^2\right)\right].
\label{coup}
\eea
The replacement $C\rightarrow C+L$ in the PST--action is required
to cope with the PST--symmetries; this replacement has to be 
made also in the transformation laws for $A^I$ in \eref{3}. The additional
term $\eta{1\over2}\int\left(L^1C^2-C^1L^2\right)$ is needed to preserve
the Dirac--anomaly. This can be seen from the the second line
of \eref{coup}: under a change of the Dirac--string the variation of the 
last term cancels the variation of $L^1dA^2-dA^1L^2$.
Since the $H^I$ are invariant under string changes
we have then $\Delta I_L=A_D=\Delta I_0$, i.e. the Dirac--anomaly in the absence
of the composite fields $L^I$. Therefore, string--independence is again
ensured by the ``old" quantization conditions.

In Schwinger--like formulations the same recipe works; one has to replace
$C\rightarrow C+L$, and to add the term 
$\eta{1\over2}\int\left(L^1C^2-C^1L^2\right)$. For the asymmetric theory 
one has to add, instead of this terms, the term $-\eta\int C^1L^2$. 
Eventually this leads for the asymmetric theory to the action 
\beq
\label{coupas}
\widetilde
I_L=I_L+{1\over 2}\eta\int{C^1C^2}.
\eeq
In the usual gauge for \eref{2}, the equations of motion for the coupled 
system become 
\bea\nonumber
T_r^2\qua^{(n)} y_r^\mu&=&\eta(-)^{nD}\,e_r^2\,{\pa y_r^{\nu_1}\over\pa\sigma^0}
\cdots  {\pa y_r^{\nu_{n+1}}\over\pa\sigma^n}\,
(F^1)^\mu{}_{\nu_{n+1}\cdots\nu_1}(y_r)\\ 
T_r^1\qua^{(D-n-4)} y_r^\mu&=&\eta(-)^{D+n+1}\,e_r^1\,
{\pa y_r^{\nu_1}\over\pa\sigma^0}
\cdots  {\pa y_r^{\nu_{D-n-3}}\over\pa\sigma^{D-n-4}}\,
(F^2)^\mu{}_{\nu_{D-n-3}\cdots\nu_1}(y_r)\nonumber\\
F^1+L^1&=&*(F^2+L^2).
\nonumber
\eea
Notice that the field strengths appearing in the Lorentz--force law
are still the  $F$'s and not the $H$'s. 
The interaction between the branes and the fields $L^I$ is thus introduced
indirectly through the modified duality relation between $F^1$ and 
$F^2$. 

\vskip0.3truecm\noindent
{\it $\vartheta$--angles for dyons.} Dyons admit generalized couplings
with $\vartheta$--angles only in $4K$--dimensional space--times. In
$D=4K+2$ one can introduce $\vartheta$--terms if one doubles the
degrees of freedom of the gauge fields. We treat here only the former
case and refer the reader for the latter case to reference \cite{DESER}.

In the symmetric theory of dyons in $D=4K$
$\vartheta$--angles are already present. This is 
due to the particular form of the relevant Dirac--Schwinger quantization 
conditions \eref{ds}, with the minus sign. Apart from an $SO(2)$--rotation, 
their general solutions can be parameterized as \footnote{For a derivation
of the solutions see \cite{LM}; w.r.t. this reference we rescaled the 
charge $e_0$ and reshuffled a factor of 2.} 
\bea\nonumber
e_r^1&=& {2\pi\over e_0} m_r\\
e_r^2&=&{n_r\over L}e_0+ \vartheta {2\pi\over e_0} m_r, 
\label{teta}
\eea
where $m_r$, $n_r$ and $L$ are integers restricted to
\beq
\label{rest}
{1\over 2}(n_rm_s-n_sm_r)=0\,\, {\rm mod}\, L.
\eeq
These solutions show up the shift of say the electric charge by an amount
proportional to the magnetic charge, which is characteristic for
$\vartheta$--angles. The integer $L$ can in general not be set
equal to $1$.

The asymmetric theory requires the Dirac--conditions \eref{d}, which have
the general solutions
\bea
\nonumber
e_r^1&=&{2\pi\over e_0} m_r\\
e_r^2&=&n_re_0,
\label{old}
\eea
and do not allow for $\vartheta$--angles. To introduce them one has to
modify the action $\widetilde I_0$ as follows
\bea
\widetilde I_0^\vartheta&\equiv& I_0[A,C^1,C^2+\vartheta C^1,a] 
                         -{1\over 2}\int C^1C^2\nonumber\\   
            &=& \int{1\over2}\left(F^1*F^1+f^1_\vartheta*f^1_\vartheta\right)
             +dA^1\left(C^2+\vartheta C^1\right)
           +{1\over2}\vartheta C^1C^1\nonumber\\         
            &=&\int{1\over2}\left(F^1*F^1+f^1_\vartheta*f^1_\vartheta\right)
             +dA^1 C^2+{\vartheta\over2}F^1F^1.
\label{asteta}
\eea
Notice that in the definition of $\widetilde I_0^\vartheta$ the shift 
$C^2\rightarrow C^2+\vartheta C^1$ has not been performed in the last term, 
because otherwise we would have obtained simply a (trivial) redefinition 
of the charges. In the second line above we used the first expression
for $I_0$ in \eref{PST}, where $f^1_\vartheta$ is obtained from $f^1$ with 
the replacement $C^2\rightarrow C^2+\vartheta C^1$. In the third line the 
$\vartheta$--term shows up in the typical way as $\int F^1F^1$ which is, 
here, not a topological term because $dF^1\neq 0$. The formulation in terms
of a single potential is obtained dropping the term 
$\int f^1_\vartheta*f^1_\vartheta$.

Since $F^1$ is string--independent, the Dirac--anomaly remains the old 
one, $\widetilde A_D$ in \eref{diracold}, the quantization conditions are
again Dirac's original ones \eref{d},
and the ``nominal" individual 
charges $e_r^I$ are still of the form \eref{old}. But since the Bianchi 
identities and equations of motion are now
\bea\nonumber 
dF^1&=&J^1\\
dF^2&=&J^2+\vartheta J^1\\
F^1&=&*F^2,
\nonumber
\eea
the physical individual charges are given by
\bea
\nonumber
E_r^1&=& {2\pi\over e_0} m_r\\
E_r^2&=& n_r e_0+ \vartheta {2\pi\over e_0} m_r,
\label{new}
\eea
where $m_r$ and $n_r$ are {\it arbitrary} integers.
These formulae have to be compared with \eref{teta}, and show once more 
the inequivalence between the symmetric and asymmetric theory. 

For what
concerns the effective action it is worthwhile to notice that the 
$\vartheta$--angle affects only the Coulomb--interaction in
\eref{gamma2}, through $C^2\rightarrow C^2+\vartheta C^1$, 
while it drops out from all electric--magnetic interactions,
the last ones being protected by the Dirac--anomaly, as shown in
\cite{LM}.

\section{Heterotic strings and five--branes in $D=10$: cancellation
         of anomalies revisited} 

The classical Green--Schwarz anomaly cancellation mechanism \cite{GS} 
applies, as it stands, to $N=1$, $D=10$ supergravity; it ensures also 
the cancellation of two--dimensional anomalies in the 
$SO(32)$--heterotic string sigma--model, in a 
supergravity background. For the $SO(32)$--heterotic five--brane, which 
still waits for a $\kappa$--symmetric sigma--model action, the Green--Schwarz
mechanism has to be amended by the so called ``inflow anomaly cancellation
mechanism" \cite{inflow}. In this section we show how this mechanism can 
be implemented through a bosonic action when a heterotic string and a 
heterotic five--brane are simultaneously present and interact with a
dynamical $N=1$, $D=10$ bosonic supergravity background. For simplicity we 
omit the dilaton and set charges and tensions to unity.

Following the conventions of the text we describe the two--dimensional 
worldvolume of the string $\Sigma_2$ by an eight--form $J^2\equiv J_8$ and the  
six--dimensional worldvolume of the fivebrane $\Sigma_6$ by a four--form 
$J^1\equiv J_4$. Dirac--strings are introduced through $J_8=dC_7$ and
$J_4=dC_3$. The corresponding supergravity potentials are conventionally 
the two-- and six--forms $A^1\equiv B_2$ and $A^2\equiv B_6$, which are 
dual to each other. 

We recall now briefly the structure of the anomalies carried by the
system. The local
symmetry groups involved are the Lorentz group $SO(1,9)$, 
the gauge group $SO(32)\otimes SU(2)$, and the structure group of
the normal bundle of the five--brane $SO(4)$; we call the corresponding
two--form curvatures respectively $R$, $F\oplus G$ and $T$. The structure
group of the normal bundle of the string $SO(8)$ does not enter the game 
since it is anomaly free. The string is neutral with respect to $SU(2)$, 
while the chiral fermions living on the fivebrane carry non trivial 
representations of $SO(32)\otimes SU(2)$. The correct field content of the
five--brane has been guessed in \cite{Witten}, and this allowed recently
to determine its total anomaly polynomial \cite{Mourad}. We report here 
the anomaly polynomials for respectively ten--dimensional supergravity 
$I_{12}$, heterotic five--branes $I_8$, and heterotic strings $I_4$
\footnote{The actual polynomials have to be multiplied by a factor of
$2\pi$, $\widehat I_n=2\pi I_n$.}:
$$
I_{12}=X_4\,X_8,\quad 
I_8=X_8+(X_4+\chi_4)\,Y_4,\quad
I_4=X_4,
$$
where
\bea
X_8&=&{1\over 192(2\pi)^4}\left(tr\,R^4 +{1\over 4}(tr\,R^2)^2
     -tr\,R^2\, tr\,F^2 +8\,tr\,F^4\right)\nonumber\\
X_4&=&{1\over 4(2\pi)^2}\left(tr\,R^2-tr\,F^2\right)\nonumber\\
Y_4&=&{1\over48 (2\pi)^2}\left(tr\,R^2-2\,tr\,T^2 -24 \,tr\,G^2\right)
\nonumber\\
\chi_4&=&{1\over 8(2\pi)^2}\,\varepsilon^{a_1a_2a_3a_4}\,
                  T^{a_1a_2}\,T^{a_3a_4}\nonumber.
\eea
For each invariant polynomial we introduce standard descent polynomials
according to 
\bea
X_8&=&dX_7\nonumber\\
\delta X_7&=&dX_6,
\eea
and we use an analogous notation for the others.
The polynomial $\chi_4$ denotes the Euler class of the five--brane normal
bundle and $a_1=(1,...,4)$. 

It should be stressed that the polynomials
formed with $R$ and $F$ are pull-backs of ten--dimensional differential
forms, i.e. forms which are defined in the whole target space, while
the ones involving $T$ and $G$ live strictly on the five--brane 
worldvolume. Therefore the string polynomial $I_4$, being a pull-back, can 
be cancelled by the ordinary Green--Schwarz mechanism while, for 
the five--brane, only the term $X_8$ in $I_8$ can be cancelled in this way.
To be more precise, an equation of the form $dH_7=I_8$ does not make sense 
since $H_7$ is a target--space form while $I_8$ is not. This is precisely
the problem faced, and solved, by the inflow mechanism.

The anomaly cancelling terms are written usually as 
\beq
I_{anom}=\left\{
\begin{array}{ll}
\int B_2X_8 &\mbox{for $D=10$ supergravity}\\
\int B_2J_8=\int_{\Sigma_2}B_2 &\mbox{for the string}\\
\int(B_6+B_2Y_4)J_4=\int_{\Sigma_6}(B_6+B_2Y_4) & \mbox{for the 
five--brane,}
\end{array}
\right.
\eeq
where the anomalous transformation laws for the potentials are 
\cite{Mourad,Witten2} $\delta B_6=-X_6$, $\delta B_2=-X_2$, but
also $\delta B_2^{(0)}=-\chi_2$ (see below for the explanation). 
Here $B_2^{(0)}$ denotes the
pull--back of $B_2$ on the five--brane, and $\chi_2$ is associated to
$\chi_4$ through the descent formalism. With these choices the above 
counterterms cancel indeed the anomalies associated through the
descent formalism to $I_{12}$, $I_4$ and $I_8$ respectively.
However, 
the last transformation law presents a problematic feature: while in the 
target space $B_2$ transforms by $-X_2$, when restricted to the five--brane
it should transform also by $-\chi_2$. But in the sense of distributions,
from the point of view of the target space field $B_2$, the 
last transformation is, actually, zero. On the other hand, if one considers
this transformation as non trivial, thus cancelling the five--brane 
anomaly, then the term $-\chi_2$ should affect also the variation of the
counterterm of the supergravity sector.

Without pursuing the discussion of this point, let us give now the
action which follows from our general framework and which avoids, in
particular, these
ambiguities. First of all we postulate for the invariant curvatures the 
Bianchi--identities
\bea
\nonumber
dH_3&=&J_4+X_4\\
\label{H7}
dH_7&=&J_8+X_8+Y_4J_4,
\eea
and the equation of motion
\beq
\label{dual37}
H_7=*H_3.
\eeq
The Bianchi identity for $H_3$ equals a target space form to a target
space form. The one for $H_7$ involves at the r.h.s. the eight--form 
$Y_4J_4$. $Y_4$ is a four--form which lives strictly on the five--brane,
but it is multiplied by the target--space form
$J_4$ which is essentially the $\delta$--function
on the five--brane, so the product is again a {\it target--space} 
eight--form. 

Actually, the expression $Y_4J_4$ is formal, because in 
general the product between a target--space form and a form living only
on the five--brane {\it is not} a target--space form. To be more
precise, by definition with $Z_8\equiv Y_4J_4$ we mean the target--space
eight--form whose components are
\beq
\label{esempio}
Z_{\mu_1\cdots\mu_8}(x)={1\over 2!4!}
\,\varepsilon_{\mu_1\cdots\mu_8\nu_1\nu_2}\int_{\Sigma_6} d^6\sigma 
\,\varepsilon^{i_1\cdots i_6}\pa_{i_1}y^{\nu_1}\pa_{i_2}y^{\nu_2}
Y_{i_3i_4i_5i_6}\,\delta^{10}(x-y(\sigma)),
\eeq
where the five--brane is parametrized by $y^\mu(\sigma)$, and 
$Y_{i_3i_4i_5i_6}(\sigma)$ are the components of the five--brane 
four--form $Y^4$.

The short--hand (factorized) notation $Y_4J_4$ is eventually justified 
by the important fact that the Leibnitz--rule for the differential
holds still true:
$$
d(Y_4J_4)=Y_4dJ_4+dY_4J_4.
$$
In the present case the r.h.s is zero, but this formula would hold also
if $Y_4$ were not closed and if the worldvolume of the five--brane were a 
manifold with boundary. In this more general case the two terms on the 
r.h.s. of the formula
are defined in complete analogy with \eref{esempio} and they represent
again a well defined target--space form. For this reason we continue to 
use the compact notation of \eref{H7}.

From the Bianchi identities it is now easy to reconstruct 
potentials 
\bea
H_3&=&dB_2+C_3+X_3\\
H_7&=&dB_6+C_7+X_7+Y_3J_4.
\eea
In this formula the term $Y_3J_4$ is defined in complete analogy to 
\eref{esempio}, and $Y_3$ is the Chern--Simons form 
associated to $Y_4=dY_3$.

Requiring that the curvatures are invariant under the local 
symmetries at hand, we obtain the transformation laws for the potentials:
\bea
\nonumber
\delta B_2&=&-X_2\\
\label{trans}
\delta B_6&=&-X_6-Y_2J_4,
\eea
which are now free from ambiguities since $\delta B_6$ represents a 
well--defined distribution. The additional term which
is present in the definition of $H_7$ is supported on the five--brane 
worldvolume and it is invariant under Dirac--string changes. This means
that $H_7$ and $H_3$ are invariant under string changes with the 
{\it standard} transformation laws for the potentials:
\bea
\nonumber
\Delta C_7 &=&dW_6, \quad \Delta B_6=-W_6\\
\Delta C_3 &=&dW_2, \quad \Delta B_2=-W_2.\label{pippo} 
\eea
The system of equations we got is now precisely of the form \eref{proto},
with $L^1\equiv L_3=X_3$ and $L^2\equiv L_7=X_7+Y_3J_4$. It is then 
straightforward to apply our recipe of the preceding section to write 
the corresponding action. As we said, for the case branes/dual branes
the appropriate theory is the asymmetric one. Including also the graviton 
and the kinetic terms for the branes it is given by
$$
I=\int_{\Sigma_{10}} \sqrt{G}\,R+\int_{\Sigma_2}\sqrt{g_2}+
\int_{\Sigma_6}\sqrt{g_6}+\widetilde I_L.
$$
The (asymmetric) action $\widetilde I_L$ represents the interaction between 
branes and potentials. It admits a PST--version in terms
of $B_6$ {\it and} $B_2$ (see \eref{coupas} and \eref{coup}), a 
Schwinger--like 
version in terms of only $B_2$ and one in terms of only $B_6$. The 
comparison between them is instructive, so we write them all explicitly
\footnote{For the additional overall minus sign see the note 
on page 5.}:
\bea
\widetilde I_L[B_2,B_6]&=&-{1\over 2}\int_{\Sigma_{10}}
\left[H\,P(v)\,H +(X_7+Y_3J_4+C_7)dB_2+(X_3+C_3)dB_6\nonumber\right.\\
&&\phantom{aaaaaaa}\left.+(X_7+Y_3J_4)C_3 +X_3C_7+C_3C_7\right]\nonumber
\\
\widetilde I_L[B_2]&=&-\int_{\Sigma_{10}}
\left[{1\over2}H_3*H_3 +(X_7+Y_3J_4+C_7)dB_2+(X_7+Y_3J_4)C_3\right]\nonumber
\\
\widetilde I_L[B_6]&=&-\int_{\Sigma_{10}}
\left[{1\over2}H_7*H_7 +(X_3+C_3)dB_6+X_3C_7\right].
\nonumber
\eea
Their Dirac--anomalies can be computed using \eref{pippo},
\bea
\nonumber
\Delta \widetilde I_L[B_2,B_6]&=&-\int_{\Sigma_{10}}W_2J_8\\
\nonumber
\Delta \widetilde I_L[B_2]&=&-\int_{\Sigma_{10}}W_2J_8\\
\nonumber
\Delta \widetilde I_L[B_6]&=&-\int_{\Sigma_{10}}W_6J_4,
\eea
and they are all integer. 

Eventually the form of all these actions is completely
determined by the requirements: 1) the equations of motion have to
be \eref{H7} and \eref{dual37}; 2) their Dirac--anomalies have to be 
integers (in this section we have set the charges to unity).

It remains to compute the anomalies of these actions under the local
symmetries and to check whether they cancel the quantum anomalies. 
Taking the above transformations of the potentials into account the
evaluation of the variations is a mere exercise and leads to
\bea
\nonumber
\delta\widetilde I_L[B_2,B_6]&=&-{1\over2}\int_{\Sigma_{10}}(X_2X_8+X_4X_6)
                              -\int_{\Sigma_2}X_2
-\int_{\Sigma_6}\left(X_6+{1\over2}(X_2Y_4+Y_2X_4)+Y_2J_4\right)\\
\delta\widetilde I_L[B_2]&=&-\int_{\Sigma_{10}}X_2X_8
                              -\int_{\Sigma_2}X_2
-\int_{\Sigma_6}\left(X_6+X_2Y_4 +Y_2J_4\right)\nonumber\\
\delta\widetilde I_L[B_6]&=&-\int_{\Sigma_{10}}X_4X_6
                              -\int_{\Sigma_2}X_2
-\int_{\Sigma_6}\left(X_6+ Y_2X_4+Y_2J_4\right).\nonumber
\eea
The anomaly supported on the string worldsheet corresponds in each case to
the polynomial $-I_4$. The same happens for the anomaly supported
on the target--space, corresponding to $-I_{12}$, which however comes
out in three different but cohomologically equivalent cocycles. The cocycle
produced by the PST--action is symmetrized in $X_4\leftrightarrow X_8$,
reflecting the fact that the action involves $B_2$ and $B_6$ in a symmetric
way. The anomaly supported  on the five--brane contains in each case
the term
\beq
\label{term}
-\int_{\Sigma_6}Y_2J_4=-\int_{\Sigma_{10}}Y_2J_4J_4,
\eeq
which has to be properly defined. The product of target--space forms
$J_4J_4$ is indeed ill-defined, because it contains the square of a
$\delta$--function, and amounts formally to $0\cdot\infty$. If one of the
two factors were a regular form, say $K_4$, then the other factor $J_4$
would perform simply the pull--back of $K_4$ on the five--brane
worldvolume $\Sigma_6$. If we indicate 
this pull--back form with $K^{(0)}_4$, then one has the equality
$$
J_4K_4=J_4K^{(0)}_4;
$$
since $K^{(0)}_4$ is now a form on the five--brane worldvolume the r.h.s.
of this equation is defined as in \eref{esempio}. For our original 
product this procedure would lead to the form $J^{(0)}_4$ which is,
however, again ill-defined. In the literature one performs usually the 
{\it cohomological} identification \cite{Witten2,BottTu}
$$
J^{(0)}_4\rightarrow \chi_4,
$$
where $\chi_4$ is the Euler class. 

Actually, this identification can be realized at the level of 
{\it differential forms}. For a flat metric, $G_{\mu\nu}(x)=\eta_{\mu\nu}$,
it can be realized in a very direct way as follows
\footnote{Notice that even for a flat target--space metric the normal 
curvature $T^{a_1a_2}$, composing $\chi_4$, is in general non vanishing 
due to the {\it intrinsic} curvature of the five--brane.}. 
Let us go back to the definition of $J_4$
as a four--current, more precisely to its explicit expression given in the
appendix in formula \eref{expl}, with $p=6$ and $D=10$. 
We regularize this formula by choosing a gaussian approximation for
the $\delta$--function appearing there,
$$
\delta^{10}(x-y)\rightarrow \delta^{10}_\varepsilon(x-y) \equiv
{e^{-\eta_{\mu\nu}(x-y)^\mu(x-y)^\nu /\varepsilon}\over (\pi\varepsilon)^5}.
$$
For $\varepsilon \rightarrow 0$ we have 
$\delta^{10}_\varepsilon(x-y)\rightarrow \delta^{10}(x-y)$, in the sense
of distributions. This regularization
gives rise to a smooth current $J_4^\varepsilon$,
the product $J_4J_4^\varepsilon$ is now well defined and one has the 
pull--back equality $J_4J_4^\varepsilon=J_4(J_4^\varepsilon)^{(0)}$.
The limit for $\varepsilon\rightarrow 0$ 
can now be evaluated explicitly \cite{LMT} and the result is indeed
$$
\lim_{\varepsilon \rightarrow 0}
J_4J_4^\varepsilon=\lim_{\varepsilon \rightarrow 0} J_4(J_4^\varepsilon)^{(0)}
=J_4\,\chi_4.
$$
More precisely, it can be shown that one has the local {\it pointwise}
limit on the five--brane
$$
\lim_{\varepsilon \rightarrow 0}\,(J_4^\varepsilon)^{(0)}=\chi_4.
$$
This allows to define the term in eq. \eref{term} through our limiting 
procedure as
$$
-\lim_{\varepsilon \rightarrow 0}
\int_{\Sigma_{10}}Y_2J_4J_4^\varepsilon=-\lim_{\varepsilon \rightarrow 0}
\int_{\Sigma_{10}}Y_2J_4(J_4^\varepsilon)^{(0)}=
-\int_{\Sigma_{10}}Y_2J_4\chi_4=-\int_{\Sigma_6}Y_2\,\chi_4.
$$
With this specification also the anomalies supported on the five--brane 
arise from the anomaly polynomial $-I_8$, and each of the three
actions given above cancels the quantum anomalies, a part from 
trivial cocycles. 

We remark that a more sophisticated regularization of $J_4$,
in a curved target--space, can be 
achieved by means of the theory of characteristic currents, developed
in \cite{harlaw}. This regularization enjoies the property 
that, in addition to  $\lim_{\varepsilon \rightarrow 0} J_4^\varepsilon=
J_4$, one has 
$$
(J_4^\varepsilon)^{(0)}=\chi_4,
$$
for every $\varepsilon$.

The main difference between our framework, based on a systematic 
introduction of Dirac--strings, and the usual way of presenting
the inflow mechanism, is given by the transformation laws of the 
potentials \eref{trans}, which are common to the three actions: 
$B_2$ transforms as in ordinary source--free $D=10$ supergravity,
and $\delta B_6$ carries an additional term localized at the five--brane,
which represents a well defined target--space form. The difference arises
also in the mechanism which produces the anomalies. For example, making
reference to the Schwinger--like action for $B_2$, the anomaly 
$-\int_{\Sigma_6}Y_2\,\chi_4$ is obtained in the usual 
framework, based on $\delta B_2^{(0)}=-\chi_2$, by varying the term 
$-\int_{\Sigma_{10}}Y_3J_4dB_2 =\int_{\Sigma_6}Y_4(B_2)^{(0)}$; in our
framework it is obtained by varying the term 
$-\int_{\Sigma_{10}}Y_3C_3J_4$, whose 
presence is implied by the requirement of unobservability of the 
Dirac--string.

Another peculiar feature of our framework is provided by a particular
interplay between Dirac--anomalies and gauge--anomalies.
One could cancel the heterotic
string anomaly of the classical action, $-\int_{\Sigma_2}X_2$, by adding 
the term $\int_{\Sigma_{10}} X_3C_7$; similarly one could cancel the 
$X_8$--part of
the five--brane anomaly by adding the term $\int_{\Sigma_{10}}X_7C_3$. 
But these 
counterterms would then produce non integer Dirac--anomalies.

If no heterotic string (five--brane) is present, which corresponds to 
drop $C_7$ ($C_3$), then the Schwinger--like action for $B_6$
$(B_2)$ can be written without introducing the Dirac--string $C_3$
($C_7$), since $\int_{\Sigma_{10}} C_3dB_6=-\int_{\Sigma_6}B_6$, and 
similarly for the other case. If strings and five--branes are 
simultaneously present, for Schwinger--like actions the introduction of
at least one Dirac--string ($C_3$ or $C_7$) is unavoidable, for the 
PST--action one must introduce both.

Let us finally comment on a problem which we overlooked until now, and
which concerns the term 
\beq
\label{sing}
-\int_{\Sigma_{10}}Y_3C_3J_4,
\eeq
which is present in the PST--action, with a factor of 1/2, and in the 
Schwinger--like action
for $B_2$. This term, as it stands, is again ill-defined because
the five--brane worldvolume, represented by  $J_4$, intersects its own 
Dirac--string, represented by $C_3$, on its boundary. This
leads again to a $\delta^2$--like singularity in $C_3J_4$, as happened 
for $J_4J_4$. Because of this singularity the manipulations we performed 
in computing the gauge anomalies were formal. Handling it in the same way
as the one encountered above i.e. through a gaussian regularization, one
can eventually show that the correct definition of the term is
$$
-\int_{\Sigma_{10}}Y_3\,C_3\,J_4\rightarrow -\int_{\Sigma_{10}}Y_3\,
\widetilde{\chi_3}\,J_4=-\int_{\Sigma_6}Y_3\, \widetilde{\chi_3},
$$
where $\widetilde{\chi_3}$ is an ``invariant" Chern--Simons form for
the Euler class, $d\widetilde{\chi_3}=\chi_4$. It differs from the
standard Chern--Simons form by an exact differential, 
$\widetilde{\chi_3}=\chi_3+d\Psi_2$.
With this specification
the gauge anomalies (and Dirac--anomalies) can be evaluated without
ambiguity and the results coincide with the ones stated in the text.

An explicit expression for $\widetilde{\chi_3}$ and its
geometric interpretation will be furnished in \cite{LMT}, where we will
present further applications of our framework, especially  to the $M$--theory
five--brane.

\section{Appendix}

\subsection{Proof of the invariance of the action} 

The invariance of $I_0$ under the PST--symmetries \eref{1}--\eref{3}
can be inferred easily from the form of its variation under general 
transformations of $A^I$ and $a$:
$$
\delta I_0=(-)^n\eta \int \left(
vf^1d\delta A^2+(-)^{(D+1)(n+1)}vf^2d\delta A^1+
{1\over \sqrt{-(\pa a)^2}} vf^1f^2 d\delta a \right).
$$
This formula allows also to deduce the equations of motion for $A^I$ and
$a$, respectively $d(vf^I)=0$ and 
$d\left({v\over \sqrt{-(\pa a)^2}}\,f^1f^2\right)=0$.

\subsection{Derivation of the effective action} 

Taking the gauge--fixings of the PST--symmetries into account and 
disregarding for the moment the $C$--independent normalization,
\eref{defg} is written as 
\beq
e^{i\Gamma[C]}\equiv\int \{ {\cal D} A\}_{gf}\,\{ {\cal D}a\}\,
               \,e^{iI_0[A,C,a]}\,\delta(a-a_0).
\label{gf}
\eeq
The gauge--fixings of the symmetries \eref{1} and \eref{2} have still to be
specified. 

The evaluation of the functional integral can be most easily performed if
one notes the following identity
\beq
\label{boh}
I_0[A,C,a]=\Gamma[C]+{1\over 2}\int H \Lambda(v)*H.
\eeq
$\Gamma[C]$ is defined as in \eref{gamma}, $H\equiv F^1-*F^2$, and 
$\Lambda(v)$ is an operator which sends $p^1$--forms in $p^1$--forms,
$$
\Lambda(v)=\eta(-)^{D+1}{d\delta\over\qua} +(-)^{D+n}vi_v.
$$
The functional integral over $a$ sets $v=v(a_0)\equiv v_0$ and  
the r.h.s. of \eref{gf} becomes
\beq
e^{i\Gamma[C]}\cdot\int \{ {\cal D} A\}_{gf} 
               \,e^{{i\over2}\int H \Lambda(v_0)*H}.
\label{rem}
\eeq
The $C$--dependence in the remaining functional integral is spurious and
can be eliminated through the shifts
\bea\nonumber
A^1&\rightarrow& A^1+(-)^D*{d\over\qua}(C^2-\eta*C^1)\\
A^2&\rightarrow& A^2+(-)^D\eta*{d\over\qua}(C^1-*C^2),\nonumber
\eea
which imply
$$
H\rightarrow dA^1-*dA^2.  
$$
These shifts are compatible, for example, with the linear
gauge fixings for the symmetries \eref{1},\eref{2}
$$
i_vA^I=0=d*A^I.
$$
The remaining integral in \eref{rem} gives, therefore, a 
$C$--independent constant which cancels precisely against the 
normalization. This can be seen using the identity \eref{boh}
with $C^I=0$. The result is \eref{gamma}, q.e.d.

\subsection{Integration of $p$--forms and Poincar\`e--duality}

We give here explicit coordinate representations for the Poincar\`e--dual
of a generic hypersurface. 
We normalize the integral of a generic $p$--form $\Psi_p$ on a 
$p$--dimensional hypersurface $\Sigma_p$, parameterized by 
$y^\mu(\sigma^i)$, $i=(1,\cdots,p)$, as follows
$$
\int_{\Sigma_p} \Psi_p= \int d^p\sigma\, {\pa y^{\mu_1}\over \pa\sigma^1}
\cdots {\pa y^{\mu_p}\over \pa\sigma^p}\,\Psi_{\mu_p\cdots\mu_1}.
$$
The Poincar\`e--dual to $\Sigma_p$, a $(D-p)$--form, admits then the 
coordinate representation
\beq
\Phi_{\Sigma_p}(x)={(-1)^{D+1}\over (D-p)!}\,
dx^{\mu_1}\cdots dx^{\mu_{D-p}}\,\ve_{\mu_1\cdots\mu_{D-p}\nu_1\cdots\nu_p}
\int d^p\sigma\,\delta^D(x-y(\sigma))\,
{\pa y^{\nu_1}\over \pa\sigma^1}\cdots
{\pa y^{\nu_p}\over \pa\sigma^p}.
\label{expl}
\eeq
It satisfies \eref{pd} for every $p$--form $\Psi_p$. The formula 
\eref{expl} gives explicit coordinate expressions 
for the currents $J^I$ when $\Sigma_p$ is the brane
worldvolume and the integration over the $\sigma^i$ covers the whole
worldvolume. If the worldvolume is boundaryless, as supposed in the text, 
one can introduce a $(p+1)$--dimensional hypersurface parameterized, for
example, by $y^\mu(\sigma^i,s)$, with $0\leq s \leq\infty$, such that 
$y^\mu(\sigma^i,0)=y^\mu(\sigma^i)$. Then the representations for the $C^I$
are again of the form \eref{expl}, with $p\rightarrow p+1$, where the 
integral over the new coordinate $s$ is restricted to the interval 
$[0,\infty]$.

\bigskip  
\paragraph{Acknowledgements.}

This work was supported by the 
European Commission TMR programme ERBFMPX-CT96-0045.



\end{document}